\PassOptionsToPackage{prologue,dvipsnames,table}{xcolor}

\documentclass[sigconf, nonacm]{acmart}

\newcommand\vldbdoi{XX.XX/XXX.XX}
\newcommand\vldbpages{XXX-XXX}
\newcommand\vldbvolume{18}
\newcommand\vldbissue{1}
\newcommand\vldbyear{2025}

\newcommand\vldbtitle{\shorttitle} 
\newcommand\vldbavailabilityurl{https://github.com/WASSER2545/ResQ}
\newcommand\vldbpagestyle{plain} 
\newcommand{\oursys}{\sys}
\newcommand{\sys}{\textsc{ResQ}\xspace}

\usepackage{graphicx}
\usepackage{balance}  
\usepackage{tikz}
\usepackage{subcaption}
\usepackage{xcolor}
\usepackage{cleveref}
\usepackage[ruled,linesnumbered,noend]{algorithm2e}
\usepackage{adjustbox}
\usepackage{bm}
\usepackage{array}
\newcolumntype{C}[1]{>{\centering}p{#1}}
\usepackage{booktabs}
\usepackage{multirow}
\usepackage{enumitem}
\usepackage{amsthm}
\usepackage{makecell}

\newtheorem{definition}{Definition}[section]
\newtheorem{example}{Example}[section]

\usepackage{bbold}
\usepackage{tcolorbox}
\usetikzlibrary{shapes, positioning, arrows.meta}
\usepackage{soul}
\newcommand{\stitle}[1]{\vspace{1mm}\noindent{\bf #1}}
\newcommand{\etitle}[1]{\vspace{1mm}\noindent{\underline{\em #1}}}

\usepackage{xcolor}
\usepackage{soul}

\definecolor{lightred}{RGB}{255, 200, 200}

\usepackage[noend]{algpseudocode}
\usepackage{float}
\usepackage{lipsum}
\usepackage{makecell} 


\usepackage{pgf-pie}
\usepackage{xcolor}

\definecolor{blue1}{HTML}{4285F4}
\definecolor{red1}{HTML}{EA4335}
\definecolor{yellow1}{HTML}{FBBC05}
\definecolor{green1}{HTML}{34A853}
\definecolor{gray1}{HTML}{9E9E9E}
\definecolor{orange1}{HTML}{FF6D00}

\usepackage{tikz}
\usepackage{framed}

\setcopyright{none}
\renewcommand\footnotetextcopyrightpermission[1]{} 
\begin{document}
\title{\oursys: Realistic Performance-Aware Query Generation}

\settopmatter{authorsperrow=4}
\author{Zhengle Wang}
\affiliation{%
  \institution{Purdue University}
}
\orcid{0009-0003-0221-6761}
\email{wangzhengle@cau.edu.cn}

\author{Yanfei Zhang}
\orcid{}
\affiliation{%
  \institution{Databend Cloud}
}
\email{zhangyanfei@databend.com}

\author{Chunwei Liu\texorpdfstring{$^\dagger$}{} }
\orcid{0000-0003-1481-2678}
\affiliation{%
  \institution{Purdue University}
}
\email{chunwei@purdue.edu}


\begin{abstract}

Database research and development rely heavily on realistic user workloads for benchmarking, instance optimization, migration testing, and database tuning. However, acquiring real-world SQL queries is notoriously challenging due to strict privacy regulations. While cloud database vendors have begun releasing anonymized performance traces to the research community, these traces typically provide only high-level execution statistics without the original query text or data, which is insufficient for scenarios that require actual execution. Existing tools fail to capture fine-grained performance patterns or generate runnable workloads that reproduce these public traces with both high fidelity and efficiency.
To bridge this gap, we propose \oursys , a fine-grained workload synthesis system designed to generate executable SQL workloads that faithfully match the per-query execution targets and operator distributions of production traces. \oursys constructs execution-aware query graphs, instantiates them into SQL via Bayesian Optimization-driven predicate search, and explicitly models workload repetition through reuse at both exact-query and parameterized-template levels. To ensure practical scalability, \oursys combines search-space bounding with lightweight local cost models to accelerate optimization. Experiments on public cloud traces (Snowset, Redset) and a newly released industrial trace (Bendset) demonstrate that \oursys significantly outperforms state-of-the-art baselines, achieving $96.71$\% token savings and a $86.97$\% reduction in runtime, while lowering maximum Q-error by $14.8\times$ on CPU time and $997.7\times$ on scanned bytes, and closely matching operator composition.

\end{abstract}

\maketitle

\pagestyle{\vldbpagestyle}
\begingroup\small\noindent\raggedright\textbf{PVLDB Reference Format:}\\
\vldbtitle. PVLDB, \vldbvolume(\vldbissue): \vldbpages, \vldbyear.\\
\href{https://doi.org/\vldbdoi}{doi:\vldbdoi}
\endgroup
\begingroup
\renewcommand\thefootnote{}
\footnote{
This work is licensed under the Creative Commons BY-NC-ND 4.0 International License. Visit \url{https://creativecommons.org/licenses/by-nc-nd/4.0/} to view a copy of this license. For any use beyond those covered by this license, obtain permission by emailing \href{mailto:info@vldb.org}{info@vldb.org}. Copyright is held by the owner/author(s). Publication rights licensed to the VLDB Endowment. \\
\raggedright Proceedings of the VLDB Endowment, Vol. \vldbvolume, No. \vldbissue\ %
ISSN 2150-8097. \\
\href{https://doi.org/\vldbdoi}{doi:\vldbdoi} \\
}\addtocounter{footnote}{-1}\endgroup

\ifdefempty{\vldbavailabilityurl}{}{
\vspace{.3cm}
\begingroup\small\noindent\raggedright\textbf{PVLDB Artifact Availability:}\\
The source code, data, and/or other artifacts have been made available at \url{\vldbavailabilityurl}.
\endgroup
}
\section{Introduction}
\label{sec:intro}

Modern cloud database systems rely heavily on realistic workloads to evaluate
performance, guide system tuning, and support instance optimization
(e.g., capacity planning and autoscaling)~\cite{lao2025gptuner,Misegiannis_Ritter_Giceva_München,krid2025redbench,chao2024cloud,Leis_Kuschewski_2021,zhang2025cloudybench}.
However, obtaining representative production SQL workloads is increasingly
difficult due to privacy, security, and data governance constraints.
Traditional TPC benchmarks (e.g., TPC-H/TPC-DS) remain widely used, but are
increasingly obsolete for modern cloud deployments: their fixed templates and
simplified assumptions fail to capture key production characteristics such as
skewed and long-tailed query distributions, strong repetition, and diverse operator mixes~\cite{redset}.

To better support empirical database research, several cloud
database vendors have recently released \emph{anonymized production traces},
including Snowset and Redset~\cite{snowset,redset}.
These traces contain query-level execution logs collected from real
deployments, where the original SQL text and user data are removed.
Instead, each trace record exposes partial \emph{execution targets} (e.g., CPU
time, scanned bytes, and other resource signals) and limited \emph{structural
signals} (e.g., operator composition).
While valuable, these traces are \emph{not executable} and cannot directly drive
end-to-end benchmarking, regression testing, or tuning in a controlled
environment.

This motivates a growing need for \emph{realistic performance-aware workload
synthesis}: generating executable SQL workloads whose query execution
behavior closely matches that of real user queries observed in production.
Without faithful execution profiles, synthetic workloads can lead to misleading
benchmark results, suboptimal configurations, and inaccurate performance
diagnosis~\cite{marcussurvivorship}.

\begin{figure*}
    \centering
    \includegraphics[width=0.75\linewidth]{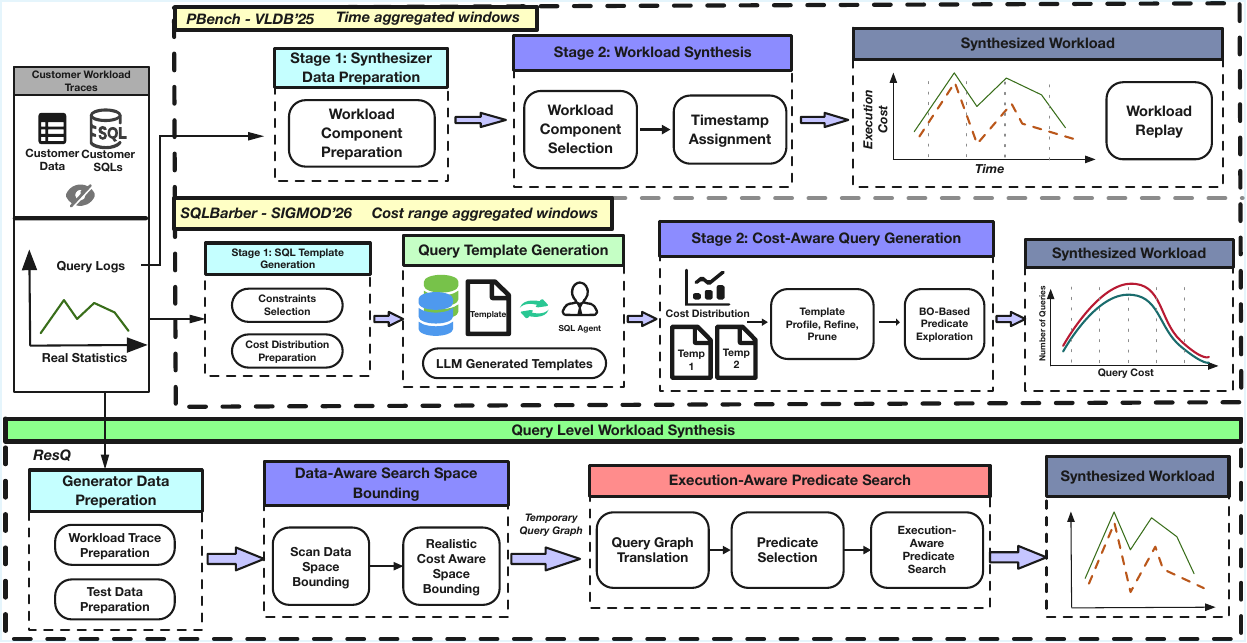}
    \vspace{-1em}
    \caption{Workload synthesis.}
    \vspace{-1em}
    \label{fig:intro}
\end{figure*}

\stitle{Coarse-Level Workload Synthesis.}
Recent work has explored workload synthesis at a coarse granularity, where
execution statistics are aggregated and modeled over time windows rather than
per individual query.
As illustrated in \Cref{fig:intro}, PBench~\cite{pbench} partitions workload traces into
consecutive time windows and synthesizes queries within each window to match
aggregated execution metrics.
PBench composes workloads by combining reusable components, obtained
from LLM-generated queries and pre-profiled benchmark fragments, to approximate
window-level execution statistics.
In contrast, SQLBarber~\cite{sqlbarber} models workload synthesis by discretizing query costs
into a fixed set of cost ranges and expanding query templates via Bayesian
Optimization to cover the target cost distribution.
To support template generation, SQLBarber further employs a
LLM-based framework to generate query templates.

However, coarse-level synthesis inherently trades fidelity for aggregation.
Because it matches targets only at a window or distribution level, it cannot
accurately capture the performance characteristics of individual queries.
Consequently, it provides limited per-query control and fails to reproduce the
strong repetition patterns prevalent in cloud workloads~\cite{redset}.


\stitle{Fine-grained Workload Synthesis.} To overcome the limitations of coarse-level synthesis, this paper proposes a \emph{fine-grained workload synthesis with realistic execution statistics}.
The goal is to generate synthetic workloads where \emph{each individual query} closely matches the execution metrics (e.g., CPU time, scanned bytes)
and structural characteristics (e.g., operator composition)
observed in real cloud workload traces. Since real workload traces do not expose the original SQL queries or the
underlying user data for privacy reasons, query-level workload synthesis
requires constructing high fidelity executable queries under incomplete information.
Specifically, this work focuses on generating synthetic queries by
(1) reconstructing performance-aware query graphs that satisfy per-query
targets and constraints derived from workload traces, and
(2) instantiating these query graphs into executable SQL queries over
public benchmark databases, such as TPC-H and TPC-DS.

\stitle{Downstream Tasks.}
Fine-grained, performance-aware synthesis enables practical workloads for
multiple downstream tasks without accessing user SQL or data:
(1) \emph{Benchmarking.} By generating query-level workloads that match real
execution profiles, the approach enables more faithful benchmarking of cloud
database systems beyond traditional template-based benchmarks.
(2) \emph{Instance Optimization.} Synthetic queries with target execution
characteristics can be used to evaluate and tune instance configurations,
such as resource allocation and scaling policies, under controlled yet
realistic workloads~\cite{ding2022sagedb,yu2024blueprinting}.
(3) \emph{Migration Testing.} During version upgrades or data migrations,
performance-aware query generation allows practitioners to replay representative
user workloads without accessing sensitive user queries or data, helping detect
performance regressions and compatibility issues~\cite{fan2024towards,li2019cloud}.
(4) \emph{Database Tuning.} By systematically varying execution targets and
data-related constraints, the generated queries support stress-testing query
optimizers and execution engines, facilitating fine-grained diagnosis and
performance tuning~\cite{lao2025gptuner,yue2024functionality}.

\stitle{Key Challenges.} Effectively addressing the problem of \emph{query-level workload synthesis with realistic execution
statistics} presents three key technical challenges.

\etitle{(C1) Fine-grained controllability under incomplete information.}
Query-level synthesis requires generating individual queries whose execution
metrics (e.g., CPU time) and operator compositions closely match those observed
in real workload traces. This demands fine-grained control over query structure
and execution behavior under incomplete information, as the original SQL
queries and databases are not available.

\etitle{(C2) Capturing workload repetition}. Production workloads often contain strong repetition, where identical or
template-similar queries recur with different parameters.
Naive synthesis tends to over-diversify queries and loses this structural and
temporal locality.

\etitle{(C3) Efficiency at scale}.
Query-level synthesis involves a large and complex search space over query
structures, predicates, and execution plans. Exhaustive exploration or
fine-grained optimization can be prohibitively expensive, while overly
simplified heuristics may lead to poor fidelity in execution statistics. Practical synthesis must balance accuracy and
generation cost.


\stitle{Our \sys Approach.} To overcome these challenges, we propose \sys (\textbf{Re}alistic \textbf{S}ynthesis of \textbf{Q}ueries). This work addresses the problem of \emph{fine-grained workload synthesis with realistic execution statistics}
by explicitly exposing and controlling the key factors that drive query execution behavior.
Rather than treating query generation as a pure black-box process, \sys adopts a structured and performance-aware
generation pipeline that directly aligns with the challenges outlined above.

First, to enable (C1), \sys adopts a
\textbf{two-phase query graph generation} framework.
In the first phase, \sys constructs abstract query graphs that define join graphs,
operator types, and high-level dataflow without committing to concrete SQL syntax.
In the second phase, these query graphs are refined with a Bayesian Optimization(BO) predicate search and instantiated into executable
queries under performance constraints. Second, to address challenge (C2), \sys deploys a \textbf{query pool}
that supports retrieval and reuse of previously generated queries at two levels:
(i) the \emph{query hash level}, which captures exact structural repetition, and
(ii) the \emph{parameterized query hash level}, which abstracts query templates by
normalizing predicate values and parameters. Finally, to address challenge (C3), \sys accelerates
the Bayesian optimization (BO) process by combining
\textbf{search space bounding} with \textbf{local predictive models}. Search space bounding algorithms are used to prune query graphs
that are \emph{provably incapable of reaching the target execution metrics}, while lightweight local prediction models provide fast cost estimates to guide BO
toward promising candidates without requiring full query executions.

\stitle{Differences from Existing Methods.}
Prior approaches either synthesize workloads at coarse granularity~\cite{pbench,sqlbarber}
or generate SQL with limited controllability using LLM prompting or
token-by-token learning~\cite{sqlstorm,Learnedsqlgen}.
These methods typically enforce execution behavior indirectly (e.g., via
prompting or rewards), making it difficult to systematically match multiple
execution targets per query and to reproduce repetition patterns.
In contrast, \sys explicitly separates \emph{structure construction},
\emph{feasibility bounding}, and \emph{predicate instantiation}, enabling
fine-grained control and practical efficiency for trace-driven query-level
synthesis. 

\stitle{Contributions.}
Our contributions are summarized as follows.

\vspace{1mm}\noindent
(1) We formalize the problem of query generation with realistic execution statistics
from anonymized cloud workload traces (\Cref{sec:preliminaries}).

\vspace{1mm}\noindent
(2) We propose \sys, a query-level synthesis framework that is both accurate and
efficient, combining repetition-aware reuse (exact-hash and parameterized-hash
query pools), performance-aware search-space bounding, and lightweight local
cost models to accelerate Bayesian-optimization-based instantiation
(\Cref{sec:framework,sec:two_phase,sec:local}).

\vspace{1mm}\noindent
(3) In collaboration with an industry cloud database vendor, we release a new
fine-grained workload trace, \textsc{Bendset}, covering diverse real-world use
cases (\Cref{sec:bendset}).

\vspace{1mm}\noindent
(4) We implement \sys and evaluate it on real cloud workload traces, including
Snowset, Redset, and \textsc{Bendset} (\Cref{sec:eval}).
\section{Problem Definition}
\label{sec:preliminaries}

\begin{definition}[\textbf{Testing Environment $\mathbf{D}$}]
We define the testing environment as the tuple $\mathbf{D}=(W,\Gamma,\Delta)$, where
$W$ denotes the target cloud database system under test (including its optimizer and execution engine),
$\Gamma$ is the collection of system-level configuration parameters (e.g., degree of parallelism, memory limits),
and $\Delta$ is the preloaded testing dataset prepared by the DBA.
The environment $\mathbf{D}$ provides a controllable sandbox for executing synthesized queries without accessing any user SQL text or user data.
\end{definition}

\begin{definition}[\textbf{Parametric Query Template}]
A \emph{parametric query template} is a SQL query skeleton $\tau$ with a parameter vector
$\bm{\theta}\in\Theta$ appearing in predicates (e.g., range bounds, equality values, \texttt{IN}-lists) and query options.
Instantiating $\tau$ with $\bm{\theta}$ yields an executable SQL query, denoted by $q=\tau(\bm{\theta})$.
\end{definition}

\begin{definition}[\textbf{Constraint Function and Constraint}]
Let $\mathcal{Q}$ denote the space of executable SQL queries in a testing environment $\mathbf{D}$.
A \emph{constraint function} is any function $f:\mathcal{Q}\times \mathbf{D}\rightarrow \mathbb{R}$ whose value can be obtained
from running $q$ in $\mathbf{D}$ and/or inspecting the resulting plan/profile (e.g., runtime, scanned bytes, operator counts).
A constraint over $f$ is specified either as a range $[l,u]$ requiring $f(q,\mathbf{D})\in[l,u]$, or as a target $t$ with tolerance
$\epsilon$ requiring $|f(q,\mathbf{D})-t|\le \epsilon$.
\end{definition}

\begin{definition}[\textbf{Targeted Query Generation}]
Given a testing environment $\mathbf{D}$, a parametric query template $\tau$ with parameter domain $\Theta$,
and a set of constraints $\mathcal{R}=\{(f_i,\rho_i)\}_{i=1}^n$ (each $\rho_i$ is a range or a target-with-tolerance),
the \emph{targeted query generation} problem is to find parameters $\bm{\theta}\in\Theta$ such that
the instantiated query $q=\tau(\bm{\theta})$ satisfies all constraints in $\mathcal{R}$ over $\mathbf{D}$.
\end{definition}

\begin{definition}[\textbf{Structure-Aware Constraints $\mathbf{C}$}]
Let $\mathcal{H}=\{h_k:\mathcal{Q}\times \mathbf{D}\rightarrow \mathbb{R}_{\ge 0}\mid k=1,\dots,K\}$ be a family of
structure-dependent functions, where each $h_k(q,\mathbf{D})$ captures a structural property of $q$ in $\mathbf{D}$
or its execution plan (e.g., number of accessed tables, number of join/aggregation/sort operators, query category).
A \emph{target structural profile} is $\mathbf{C}=(c_k)_{k=1}^K$ with tolerances $\bm{\epsilon}^H$.
A query $q$ satisfies the structure-aware constraints if $\forall k\in\{1,\dots,K\}$,
$|h_k(q,\mathbf{D})-c_k|\le \epsilon^H_k$ (exact matching is the special case $\epsilon^H_k=0$).
\end{definition}

\begin{example}[\textbf{Structure-aware constraints}]
In traces such as Snowset, Redset, and \textsc{Bendset}, the original SQL is not available, but a record may specify
a structural profile, e.g., ``tables accessed $=3$'', ``\#joins $=2$'', and ``\#sorts $=1$''.
Given a testing environment $\mathbf{D}$ over a proxy dataset $\Delta$, structure-aware constraints require synthesized queries
to match these structural signals (within tolerances).
\end{example}

\begin{definition}[\textbf{Execution Performance Targets $\mathbf{Y}$}]
Let $\mathcal{G}=\{g_m:\mathcal{Q}\times \mathbf{D}\rightarrow \mathbb{R}_{\ge 0}\mid m=1,\dots,M\}$ be a family of measurable
execution-performance functions, where each $g_m(q,\mathbf{D})$ is an observed physical execution metric of $q$ in $\mathbf{D}$
(e.g., CPU time, I/O time, scanned bytes, peak memory, elapsed time).
A \emph{target execution profile} is a vector $\mathbf{Y}=(y_m)_{m=1}^M$ with tolerances $\bm{\epsilon}^G$.
A query $q$ matches $\mathbf{Y}$ if $\forall m$, $|g_m(q,\mathbf{D})-y_m|\le \epsilon^G_m$.
\end{definition}

\begin{example}[\textbf{Execution targets}]
Consider a testing environment $\mathbf{D}$ corresponding to a specific cloud database version and configuration.
A trace record may report targets such as: CPU time $=2.4$s,
peak memory $=512$MB, and scanned bytes $=800$MB. These values form the target execution profile $\mathbf{Y}$.
\end{example}

\begin{definition}[\textbf{Realistic Performance-Aware Query Generation (RPQG)}]
Given (i) a testing environment $\mathbf{D}$, (ii) a target structural profile $\mathbf{C}$,
(iii) a target execution profile $\mathbf{Y}$, and (iv) a search space of parametric templates $\mathcal{T}$,
the \emph{realistic performance-aware query generation} problem is to synthesize an executable query
$q^\ast=\tau(\bm{\theta})$ for some $\tau\in\mathcal{T}$ and $\bm{\theta}\in\Theta_\tau$ such that:
(a) $q^\ast$ satisfies the structure-aware constraints $\mathbf{C}$, and
(b) $q^\ast$ matches the execution targets $\mathbf{Y}$ as closely as possible when executed in $\mathbf{D}$.

When exact satisfaction is infeasible, RPQG minimizes a mismatch objective over execution metrics, e.g.,
\[
\min_{q=\tau(\bm{\theta})}\;\sum_{m=1}^M w_m\cdot
\frac{|g_m(q,\mathbf{D})-y_m|}{\max(y_m,\eta)}
\quad \text{s.t.}\quad |h_k(q,\mathbf{D})-c_k|\le \epsilon^H_k,\ \forall k,
\]
where $w_m$ are user-specified weights and $\eta>0$ avoids division by 0.
\end{definition}
\section{System Architecture of \sys}
\label{sec:framework}

\begin{figure*}[h]
    \centering
    \includegraphics[width=0.75\linewidth]{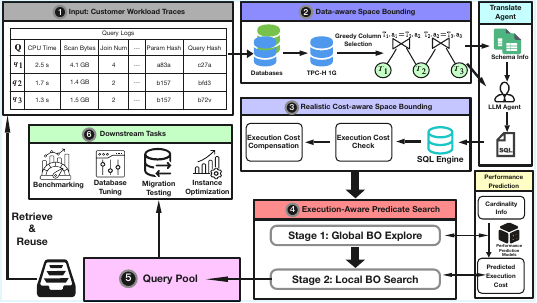}
    \vspace{-1em}
    \caption{System architecture of \sys. }
    \vspace{-1em}
    \label{fig:arch}
\end{figure*}

This section describes the architecture of \sys for solving the
\emph{realistic performance-aware query generation} problem defined in
\Cref{sec:preliminaries}.
Given an anonymized workload trace and a privacy-preserving testing environment
$\mathbf{D}=(W,\Gamma,\Delta)$, \sys synthesizes executable SQL queries that
match (within tolerances) the trace-provided \emph{structure-aware constraints}
$\mathbf{C}$ and \emph{execution performance targets} $\mathbf{Y}$, without
access to the original user SQL or data.

As shown in \Cref{fig:arch}, \sys takes as input a sequence of anonymized
query-level trace records, where each record specifies (i) a set of execution
targets (e.g., CPU time and scanned bytes) and (ii) a set of structure-aware
constraints (e.g., join/aggregation/sort composition) (\Cref{sec:preliminaries}).
For each record, \sys synthesizes an executable SQL query in the testing
environment $\mathbf{D}$ that satisfies the structural constraints and matches
the execution targets as closely as possible.

To make this feasible at scale, \sys combines: (i) generator data preparation
; (ii) a two-phase generation pipeline that first bounds
the data and performance search space and then instantiates predicates
(\Cref{sec:two_phase}); (iii) a query graph translator that converts the
execution-oriented representation into executable SQL;
(iv) local performance models to reduce expensive query executions during search
(\Cref{sec:local}); and (v) a query pool that enables retrieval and reuse to
recover repetition and improve generation efficiency.

\subsection{Generator Data Preparation}
\label{sec:preprocess}
This stage comprises two key processes: workload trace preparation and test data
preparation.

\subsubsection{\textbf{Workload Trace Preparation}}
Popular traces released by cloud DB vendors provide low-sensitivity
performance signals that can be shared for research while omitting the original
SQL text and user data. Following \Cref{sec:preliminaries}, we view each trace
record as specifying (a) an execution target vector $\mathbf{Y}$ and (b) a
structure-aware profile $\mathbf{C}$, along with its timestamp.
Unlike window-level synthesis (e.g., PBench~\cite{pbench}) that aggregates
query-level records into coarse time windows, \sys preserves query-level records
to enable fine-grained target matching and recovery of repetition patterns.

\textbf{Execution targets ($\mathbf{Y}$).}
In our current implementation, we extract per-query \emph{CPU time} and
\emph{scanned bytes} from Bendset, Snowset~\cite{snowset}, and
Redset~\cite{redset} as the primary execution targets.

\textbf{Structure-aware constraints ($\mathbf{C}$).}
We constrain operator composition using \emph{join}, \emph{aggregation}, and
\emph{sort} as the key operator types.
Bendset and Redset report explicit operator counts, while the earliest public
trace Snowset only provides coarse operator-level profiled time; therefore, for
Snowset we treat operator occurrence as a binary indicator (present/absent).
More details of Bendset are introduced in \Cref{sec:bendset}.

\subsubsection{\textbf{Test Data Preparation}}
To generate executable queries without accessing user data, we instantiate the
testing environment $\mathbf{D}=(W,\Gamma,\Delta)$ (\Cref{sec:preliminaries})
using privacy-preserving proxy datasets $\Delta$.
We deploy multiple instances of TPC-H and TPC-DS with different scale factors as
candidate datasets. Dataset scale primarily impacts scanned bytes, while data
distributions (e.g., skew) influence join fan-out and intermediate result sizes,
thereby affecting CPU time.

To support subsequent generation stages, we preprocess each proxy dataset
$\Delta$ at both the schema and data-distribution levels.
For schema metadata, we record table/column names, data types, and join keys.
For data-distribution metadata, we collect row counts, per-column data sizes,
and basic value statistics (e.g., min/max). These metadata are used later to
construct feasible query structures, bound scanned bytes, and define predicate
domains during predicate search.

\subsection{Two-Phase Query Graph Generation}
\label{sec:arch_two_phase}
As illustrated in \Cref{fig:arch}, \sys generates each query using a two-phase
pipeline (detailed in \Cref{sec:two_phase}).
The key idea is to first bound the search space using constraints that are
largely determined by \emph{structure} and \emph{data access}, and then refine
the remaining degrees of freedom using \emph{predicate instantiation} to match
execution targets.

\textbf{Phase I: Space bounding and base graph construction.}
\sys first builds a base query graph that satisfies the structure-aware
constraints $\mathbf{C}$ and is feasible over the proxy dataset $\Delta$.
This phase performs \emph{data-aware space bounding} to prune candidates that
cannot reach the scanned-bytes target in $\mathbf{Y}$, and produces a small set
of feasible base graphs for refinement.

\textbf{Phase II: Cost-aware refinement and predicate instantiation.}
Starting from a bounded base graph, \sys refines the query toward the target CPU
time in $\mathbf{Y}$.
It combines realistic cost-aware checks (and, when needed, lightweight cost
compensation) with an performance-aware predicate search procedure to tune
selectivities efficiently.

\subsection{\textbf{Query Graph Translator}}
\label{sec:LLM_trans}
To bridge the gap between operator-level planning and executable SQL, \sys
deploys a specialized query graph translator.
This component does not participate in cost-aware decision making. Instead, it
operates on a fully specified query graph and is responsible for producing a
runnable SQL query in the dialect of $W$.

The translator deterministically maps relational operators (e.g., Scan, Join,
Aggregate, Sort, EvalScalar) to their SQL counterparts while preserving operator
dependencies. An LLM-based module is used only to resolve syntactic and
dialect-specific details that are difficult to express with deterministic rules,
such as subquery nesting, alias management, and expression placement.

\subsection{Performance Prediction}
\label{sec:arch_prediction}
Performance-aware generation requires frequent feedback about execution metrics,
but running every candidate query in $W$ can be prohibitively expensive. \sys
therefore incorporates lightweight local performance models to estimate key
metrics---especially CPU time---from the query graph and available statistics.
These estimates are used to guide cost-aware bounding and to reduce the number
of expensive executions during predicate search.
The detailed model design is described in \Cref{sec:local}.

\subsection{Query Pool}
\label{sec:query_pool}
Real-world workloads exhibit strong repetition. To recover repetition patterns
and reduce generation latency, \sys maintains a query pool that supports
retrieval and reuse at two levels: (i) \emph{query-hash} reuse for exact
structural repetition, and (ii) \emph{parameterized-hash} reuse for template-level
repetition where the structure is reused but predicate values differ.
When a trace record matches an entry in the pool, \sys reuses the corresponding
structure/template and focuses computation on lightweight refinement and
predicate instantiation, improving efficiency while preserving workload
locality.

In practice, some workload traces (e.g., Snowset) do not expose explicit query
hashes due to privacy constraints. In such cases, \sys constructs a
\emph{proxy signature} by combining the target execution profile $\mathbf{Y}$
and structural constraints $\mathbf{C}$ into a fixed-length feature vector.
This vector is used as an index to retrieve candidate queries or query graphs
from the pool whose execution behavior and structure best match the given
trace record. If a suitable match is found, \sys directly reuses the retrieved
query (or query graph); otherwise, the record is forwarded to the generation
pipeline to synthesize a new query, which is then inserted into the pool for
future reuse.
\section{Two-Phase Query Graph Generation}
\label{sec:two_phase}
This section describes the core generation pipeline of \sys.
Given a trace record that specifies structure-aware constraints $\mathbf{C}$ and
execution targets $\mathbf{Y}$ (\Cref{sec:preliminaries}), our goal is to
synthesize an executable query that (i) satisfies $\mathbf{C}$ and (ii) matches
$\mathbf{Y}$ as closely as possible when executed in the testing environment
$\mathbf{D}=(W,\Gamma,\Delta)$.

The main difficulty is that $\mathbf{Y}$ contains \emph{execution-level} signals
(e.g., CPU time and scanned bytes), while directly searching in the space of SQL
syntax is both expensive and poorly controllable. \sys addresses this by
generating an execution-oriented representation first, and only translating to
SQL after the structure and cost feasibility are largely determined.

\subsection{From Realistic Query Generation to Query Graph Generation}
\subsubsection{\textbf{Why Query Graph}}
Execution targets such as CPU time and scanned bytes are primarily driven by
operator composition (e.g., number of joins, aggregations, sorts), accessed data
(volume and width), and predicate selectivities---not by superficial SQL syntax.
Therefore, instead of generating SQL text directly, \sys introduces \emph{query graph} $\mathbf{G}$, an abstraction that captures operator dependencies and
dataflow while leaving dialect-specific syntax and physical implementation
details to later stages.

\subsubsection{\textbf{Query Graph}}
\begin{definition}[\textbf{Query Graph $\mathbf{G}$}]
A \emph{query graph} $\mathbf{G}=(\mathcal{V},E)$ is a directed acyclic graph
(DAG) whose nodes $\mathcal{V}$ represent relational operators (e.g., Scan, Join,
Aggregate, Sort, EvalScalar) and whose edges $E$ represent data dependencies
between operators. Each node carries attributes such as referenced tables,
columns, and operator parameters needed for SQL instantiation.
\end{definition}

\subsubsection{\textbf{Problem Reformulation}}
The RPQG objective in \Cref{sec:preliminaries} is defined over executable queries
$q$ via structure functions $h_k(q,\mathbf{D})$ and execution functions
$g_m(q,\mathbf{D})$. \sys uses $\mathbf{G}$ as an intermediate representation
and splits generation into:
(i) constructing feasible $\mathbf{G}$ that satisfy the structural profile
$\mathbf{C}$ and can reach data-access targets (e.g., scanned bytes), and
(ii) instantiating $\mathbf{G}$ into a parametric SQL template and tuning
predicates to match compute targets.

Conceptually, this reduces RPQG to a constrained search over query graphs, plus a
parameter search within each graph.

\begin{definition}[\textbf{Multi-Target Query Graph Generation}]
Let $\mathbb{G}$ be the space of valid query graphs over the schema of $\Delta$.
Given a target structural profile $\mathbf{C}$ and a target execution profile
$\mathbf{Y}$, the multi-target query graph generation problem is to find a query
graph $\mathbf{G}^\ast\in\mathbb{G}$ such that $\mathbf{G}^\ast$ satisfies
$\mathbf{C}$ and admits an instantiation whose execution in $\mathbf{D}$
matches $\mathbf{Y}$.
\end{definition}

\begin{proposition}[\textbf{Query-to-Graph Reduction (Informal)}]
For the execution targets considered in this paper (e.g., scanned bytes and CPU
time), feasibility and optimization are largely determined by (i) operator
composition and dependencies and (ii) accessed data volume and predicate
selectivity. These factors are explicitly represented in a query graph.
Therefore, \sys first searches in the space of query graphs to control these
factors, and then translates the chosen graph into SQL for final instantiation.
\end{proposition}

\subsection{Phase I: Data-Aware Search Space Bounding}
\label{sec:data_aware_bounding}
Phase~I bounds the search space using \emph{data-access feasibility} under the
proxy dataset $\Delta$. In particular, we target scanned bytes, since it is
strongly affected by the choice of scanned tables/columns and can be bounded
before predicate tuning.

\subsubsection{\textbf{Scan-Bytes Space Bounding}}
\label{sec:scan-bytes-bounding}
Let $\mathcal{R}=\{R_1,\dots,R_n\}$ denote the tables in $\Delta$.
Given a structural constraint on joins (e.g., join count in $\mathbf{C}$), a
candidate query graph must select a set of base scan operators, which induces a
set of accessed tables $\mathcal{R}'\subseteq\mathcal{R}$.
To ensure query validity, $\mathcal{R}'$ must form a connected subgraph in the
schema join graph (e.g., induced by PK--FK relationships).

For each selected table $R_i\in\mathcal{R}'$, let $\mathcal{A}^{(i)}$ be its
columns. Each column $A^{(i)}_j$ is associated with an estimated scan weight
$w^{(i)}_j$ (bytes) derived from storage statistics collected during
\Cref{sec:preprocess}. Given a target scanned-bytes value (from $\mathbf{Y}$),
the goal is to pick a subset of columns per table such that the estimated scan
bytes matches the target as closely as possible, while ensuring join-key columns
are included for feasibility.

This yields a constrained multi-subset selection problem that we use as a
\emph{bounding step}: graphs that cannot reach the scan target are pruned early,
before any expensive predicate optimization.

\subsubsection{\textbf{Greedy Data-Aware Column Selection}}
\label{sec:greedy_scan}
An exact dynamic programming solution is impractical for realistic schemas due
to exponential complexity in the number of columns. \sys therefore adopts a
greedy algorithm (Algorithm~\ref{alg:greedy-scan}) that scales to iterative
generation. The algorithm proceeds in three phases:
(i) \emph{join graph grounding} to select a connected table set and mandatory
join-key columns, (ii) \emph{per-table initialization} to avoid degenerate scans
(each table contributes at least one column), and (iii) \emph{global greedy
completion} that adds additional columns until the scan target is reached.

\textbf{Complexity and reuse.}
For a fixed join graph, the greedy column selection runs in
$O(N\log N)$ time, where $N$ is the number of candidate columns across the
selected tables. Although enumerating join graphs is exponential in the worst
case, the number of feasible connected subgraphs is small for structured
benchmark schemas (e.g., TPC-H/TPC-DS). Moreover, \sys caches intermediate
results---grounded join graphs and their scan-bounded column selections---and
reuses them across trace records with similar structural profiles, substantially
reducing repeated work.

\subsection{Phase I: Realistic Cost-Aware Space Bounding}
\label{sec:cost_aware_bounding}
After bounding scanned bytes, \sys further refines candidate query graphs using
compute-related feasibility, with a focus on CPU time.

\begin{algorithm}[t]
\caption{Greedy Data-Aware Column Selection}
\label{alg:greedy-scan}
\KwIn{
Target scanned bytes $y_{\text{scan}}$;
Schema-level join graph $\mathcal{G}_{\text{schema}}$;
Candidate table set $\mathcal{R}$ with per-table column sets
$\{\mathcal{A}^{(i)}\}$
}
\KwOut{
Column selections $\{\mathcal{S}^{(i)}\}_{i=1}^{K}$
}

\tcp{\textbf{Step I: Join Graph Grounding}}
Select a subset $\mathcal{R}' \subseteq \mathcal{R}$ such that
$|\mathcal{R}'| = K$ and
$\mathcal{R}'$ forms a connected subgraph in
$\mathcal{G}_{\text{schema}}$\;
Derive table groups
$\{\mathcal{A}^{(i)}\}_{i=1}^{K}$ from $\mathcal{R}'$\;
Identify mandatory join-key columns
$\{\mathcal{M}^{(i)}\}_{i=1}^{K}$ from join predicates\;

Initialize $\mathcal{S}^{(i)} \leftarrow \mathcal{M}^{(i)}$ for all $i$\;
$S \leftarrow \sum_{i} \sum_{A^{(i)}_j \in \mathcal{M}^{(i)}} w^{(i)}_j$\;

\tcp{\textbf{Step II: Per-Group Initialization}}
Set average target $\bar{y} \leftarrow y_{\text{scan}} / K$\;
\For{$i \gets 1$ to $K$}{
    \If{$\mathcal{S}^{(i)} = \emptyset$}{
        Select
        $A^{(i)}_{j^*}
        = \arg\min_{A^{(i)}_j \in \mathcal{A}^{(i)}}
        |w^{(i)}_j - \bar{y}|$\;
        $\mathcal{S}^{(i)} \leftarrow \{A^{(i)}_{j^*}\}$\;
        $S \leftarrow S + w^{(i)}_{j^*}$\;
    }
}

\tcp{\textbf{Step III: Global Greedy Completion}}
Construct candidate set
$\mathcal{R}_{\text{cand}}
= \{ A^{(i)}_j \mid
A^{(i)}_j \in \mathcal{A}^{(i)} \setminus \mathcal{S}^{(i)} \}$\;
Sort $\mathcal{R}_{\text{cand}}$ by ascending $w^{(i)}_j$\;

\ForEach{$A^{(i)}_j \in \mathcal{R}_{\text{cand}}$}{
    \If{$S \ge y_{\text{scan}}$}{
        \textbf{break}
    }
    $\mathcal{S}^{(i)} \leftarrow \mathcal{S}^{(i)} \cup \{A^{(i)}_j\}$\;
    $S \leftarrow S + w^{(i)}_j$\;
}

\Return $\{\mathcal{S}^{(i)}\}_{i=1}^{K}$\;
\end{algorithm}

\textbf{Step I: enforce structure-aware constraints.}
Starting from a scan-bounded base graph, \sys injects additional operators to
satisfy the target structural profile $\mathbf{C}$ (e.g., adding Sort and
Aggregate operators when required), producing a complete candidate graph.

\textbf{Step II: execution-cost feasibility check.}
We obtain an initial CPU-time estimate for the candidate graph and compare it
with the target CPU time from $\mathbf{Y}$. This check uses local performance
models (\Cref{sec:local}) whenever possible. 
This design provides a practical trade-off between
accuracy and cost, avoiding full executions for most candidates.

\textbf{Step III: execution cost compensation.}
If the estimated CPU time is below the target, \sys introduces lightweight
\texttt{EvalScalar} operators that apply computational expressions over selected
columns to increase CPU cost while keeping the overall query structure
unchanged. Candidate scalar operations are chosen from a supported library
(arithmetic, string, date functions). Their incremental CPU cost is estimated by
local models (\Cref{subsec:local_models}), and \sys greedily selects operation
types and application counts to bring the query graph into a feasible range.
This ensures that the subsequent predicate search has enough ``room'' for
fine-grained adjustment toward the final target.

\subsection{Phase II: Performance-Aware Predicate Search}
\label{sec:predicate_search}
Phase~II instantiates predicates to match execution targets, focusing on CPU
time. Given a fixed query structure (query graph) and a set of candidate
predicate columns, \sys searches for predicate parameters that yield CPU time as
close to the target as possible while minimizing expensive executions.




\subsubsection{\textbf{Predicate Column Selection}}
\label{sec:predicate-selection}
Allowing all possible predicate columns leads to a high-dimensional and
irregular search space, which significantly degrades Bayesian optimization (BO).
\sys therefore restricts tuning to a small set of \emph{execution-sensitive}
columns. Intuitively, CPU time is strongly influenced by intermediate result
sizes; hence, predicates that (i) filter early and (ii) offer a wide and smooth
selectivity range are the most useful knobs.

For multi-join queries, \sys does not require each table to contribute the same
number of predicate dimensions. Instead, more dimensions are allocated to large
tables or tables with wider value ranges, since they typically dominate CPU-time
variance. We restrict predicate forms to single-sided range predicates (e.g.,
$R.a \le p$), which yields a weak monotonic relationship between predicate values
and intermediate result sizes, improving optimization stability.

\subsubsection{\textbf{Two-Stage Predicate Tuning}}
Predicate tuning is formulated as black-box optimization over a bounded domain
constructed from the selected predicate columns. \sys adopts a two-stage BO
strategy: Stage~1 performs global exploration to locate promising regions, and
Stage~2 performs local refinement around representative seeds near the target.

\subsubsection{\textbf{Performance-Aware Scoring Function}}
\label{sec:execution-aware-score}
Directly executing the query for every candidate predicate configuration is
expensive. \sys therefore uses a hybrid scoring strategy: it first predicts CPU
time using local models, and only executes the query when the predicted CPU time
falls within a tolerance window around the target. This preserves accurate
feedback near the target region while dramatically reducing executions far from
the target.

\subsubsection{\textbf{Stage 1: Global BO Exploration}}
In Stage~1, BO is initialized with a small number of random evaluations,
followed by guided exploration. All evaluated predicate configurations and their
scores are recorded. At the end of Stage~1, \sys selects two seeds:
$x^{+}$ (closest to but not smaller than the target) and $x^{-}$ (closest to but
smaller than the target), which bracket the target region from both sides.

\subsubsection{\textbf{Stage 2: Local BO Refinement}}
In Stage~2, \sys constructs reduced search spaces by shrinking each predicate
dimension around $x^{+}$ and $x^{-}$, and optionally applies value bucketing for
very large domains. BO is then run on each reduced space with an emphasis on
exploitation. Refining from both sides improves robustness against prediction
errors and irregular cost surfaces.

Algorithm~\ref{alg:predicate_bo} summarizes the two-stage tuning
procedure.

\begin{algorithm}[h]
\caption{Two-stage BO Predicate Tuning}
\label{alg:predicate_bo}
\KwIn{
    SQL template $q^{temp}$, predicate columns $C$, column types $T$, 
    target CPU time $y^{tar}_{cpu}$
}
\KwOut{Optimized SQL predicates}

\SetAlgoLined

\BlankLine
\tcp{\textbf{Helper function: Score Predicates}}

\SetKwFunction{FMain}{ScorePredicate}
\SetKwProg{Fn}{Function}{:}{}
\Fn{\FMain{$x$}}{
    $q' \gets \textsc{BuildSql}(x, q^{temp})$\
    $y^{pred}_{cpu} \gets \textsc{localmodelpredict}(q')$\;
    \eIf{$y^{pred}_{cpu} \in [y^{tar}_{cpu}\cdot a, y^{tar}_{cpu}\cdot b]$}{
        $y^{obs}_{cpu} \gets \textsc{RunQuery}(q')$
    }{
        $y^{obs}_{cpu} \gets y^{pred}_{cpu}$
    }
    $score \gets -\mathbf{MSE}\left(y^{obs}_{cpu}, y^{tar}_{cpu}\right)$\;
    \KwRet $\text{score}$\;
}

\BlankLine
\tcp{\textbf{Stage 1: Global BO Explore}}

$\mathcal{S}_1 \gets \textsc{BuildGlobalSearchSpace}(C, T)$\;
Initialize BO on $\mathcal{S}_1$\;

\For{$i=1$ \textbf{to} $n_{\text{rand}} + n_{\text{calls}}$}{
    $x \gets \text{BO.ask()}$\;
    $\text{score} \gets$ \FMain{$x$}\;
    $\text{BO.tell}(x, \text{score})$\;
    Record $x$, score\;
}
Select seeds for Stage 2: $x^{+}$ and $x^{-}$\;

\BlankLine
\tcp{\textbf{Stage 2: Local BO Search}}

\For{each seed $x^{*} \in \{x^{+}, x^{-}\}$}{
    $\mathcal{S}_2 \gets \textsc{ShrinkSearchSpace}(\mathcal{S}_1, x^{*})$\;
    $\mathcal{S}_2 \gets \textsc{ApplyBucketing}(\mathcal{S}_2)$\;
    Initialize BO on $\mathcal{S}_2$\;
    
    \For{$i=1$ \textbf{to} $n_{\text{rand}} + n_{\text{calls}}$}{
        $x \gets \text{BO.ask()}$\;
        $\text{score} \gets$ \FMain{$x$}\;
        $\text{BO.tell}(x, \text{score})$\;
    }
}

\BlankLine
\Return $\{x^{+}_{best}, x^{-}_{best}\}$
\end{algorithm}

\section{Local Model Support}
\label{sec:local}
Constraint-aware generation in \sys requires frequent feedback on execution
targets $\mathbf{Y}$ (e.g., CPU time) defined in \Cref{sec:preliminaries}.
While executing each candidate query in the testing environment
$\mathbf{D}=(W,\Gamma,\Delta)$ yields accurate measurements, doing so inside the
search loop is prohibitively expensive.
\sys therefore uses \emph{local performance models} that approximate key
execution metrics---most importantly CPU time---from operator-level features of
a candidate query graph and its intermediate statistics.
These models are ``local'' in the sense that they are trained and used under a
fixed environment $\mathbf{D}$ (i.e., a specific engine $W$, configuration
$\Gamma$, and dataset $\Delta$), which substantially improves prediction
stability.

This section presents our CPU-time modeling and prediction framework.
We model execution at the granularity of physical operators and aggregate
operator-wise predictions to estimate the query-level CPU time. We begin with
join operators, since they are typically the dominant CPU consumers in
analytical workloads and are the hardest to model accurately with simple
analytical formulas. We then describe lightweight models for other operators.

\subsection{Modeling of Join Operators}
\label{sec:join_model}
Across different join algorithms, CPU consumption is largely dominated by two
common phases:
(i) a \emph{matching phase}, which processes join keys and identifies matching
tuple pairs, and
(ii) a \emph{materialization phase}, which constructs output tuples and
materializes selected columns.
We therefore model the CPU time of a join operator as
\begin{equation}
\label{eq:join_cpu_decomp}
CPU_{\text{join}} \;=\; CPU_{\text{match}}^{(\cdot)} \;+\; CPU_{\text{material}},
\end{equation}
where $CPU_{\text{match}}^{(\cdot)}$ captures algorithm-specific matching logic,
and $CPU_{\text{material}}$ captures output construction shared across join
algorithms.

\subsubsection{\textbf{Hash Join}}
\label{sec:hj_cpu}
Modern vectorized OLAP engines commonly implement hash join as a matching phase
followed by an optional materialization phase. During matching, the operator
produces lightweight row references (e.g., row indices) rather than materialized
tuples. During materialization, selected columns are fetched/decoded based on
these references.

\stitle{Matching Phase.}
The matching phase computes hash values for join keys and probes the hash table.
Its CPU cost is primarily driven by input cardinalities and per-tuple hashing
cost:
\begin{equation}
\label{eq:hj_match}
CPU_{\text{match}}
=
\alpha \cdot |R_B| \cdot cost_{\text{hash}}(K_B)
+ \beta \cdot |R_P| \cdot cost_{\text{hash}}(K_P)
+ \delta \cdot \#\text{tasks},
\end{equation}
where $|R_B|$ and $|R_P|$ are build/probe cardinalities, $K_B$ and $K_P$ are join
keys, and $\#\text{tasks}$ captures orchestration overhead under the execution
environment (e.g., parallel fragments).

\stitle{Materialization Phase.}
Materialization fetches/decodes output columns according to the row references
and is driven by output cardinality and projected width:
\begin{equation}
\label{eq:hj_take}
CPU_{\text{material}}
\approx
\epsilon \cdot |R_{\text{out}}| \cdot cost_{\text{fetch}}(S_{\text{out}}),
\end{equation}
where $|R_{\text{out}}|$ is output cardinality, $S_{\text{out}}$ is the output
schema, and $cost_{\text{fetch}}(\cdot)$ captures per-row fetch/decoding cost
that depends on column count and data types.

\stitle{Overall Hash Join CPU Time.}
Combining the two phases,
\begin{equation}
\label{eq:hj_total}
CPU_{\text{HJ}} = CPU_{\text{match}} + CPU_{\text{material}}.
\end{equation}
This decomposition explains why CPU time can vary substantially between
count-only joins (minimal materialization) and joins that project wide tuples
(materialization-dominated).

\subsubsection{\textbf{Other Join Algorithms}}
For nested-loop join and merge join, we adopt the same two-phase abstraction in
\Cref{eq:join_cpu_decomp}. Their matching phases differ in implementation (e.g.,
nested probing vs.\ sequential comparison), but their materialization costs are
captured using the same width-aware modeling strategy.
Since hash join dominates analytical workloads in our target systems, we focus
our learning effort on hash joins and treat other algorithms as secondary cases.

\subsection{Modeling Other Operators}
\label{subsec:local_models}
Beyond joins, many operators exhibit simpler CPU-time patterns in modern OLAP
engines and can be approximated with lightweight parametric models.

\stitle{Scan.}
Scan CPU time is driven by the number of input tuples and the cost of
decoding/processing accessed column segments:
\begin{equation}
\label{eq:cpu_scan}
CPU_{\text{scan}}
=
\alpha_s \cdot |R| \cdot \sum_{c \in \text{cols}} S_{\text{segment}}(c),
\end{equation}
where $|R|$ is the number of scanned tuples, \text{cols} is the accessed column
set, and $S_{\text{segment}}(c)$ is an average per-tuple segment-processing cost
(e.g., decompression/decoding) derived from local profiling.

\stitle{Projection and Scalar Functions.}
For projection and scalar computations (e.g., string/date functions), CPU time is
approximately linear in input cardinality:
\begin{equation}
\label{eq:cpu_func}
CPU_{\text{func}} = \alpha_g \cdot |R| \cdot cost_{\text{func}}.
\end{equation}

\stitle{Sort.}
For comparison-based sorting, CPU time typically follows a super-linear trend:
\begin{equation}
\label{eq:cpu_sort}
CPU_{\text{sort}} = \alpha_o \cdot |R| \cdot \log |R|.
\end{equation}

These abstractions are not meant to be perfect analytical models; rather, they
provide compact features/priors that enable efficient learning and stable
extrapolation in a fixed environment $\mathbf{D}$.

\subsection{CPU Time Prediction}
\label{sec:cpu_prediction}
\sys estimates the query-level CPU time by aggregating operator-level
predictions. Given a candidate query graph $\mathbf{G}$ with operator nodes
$\mathcal{V}$, we compute:
\begin{equation}
\label{eq:cpu_sum}
\widehat{CPU}(\mathbf{G})
=
\sum_{v \in \mathcal{V}} \widehat{CPU}(v),
\end{equation}
where $\widehat{CPU}(v)$ is the predicted CPU time for operator $v$.

We distinguish between \emph{complex operators} (primarily joins), whose CPU time
depends on multiple interacting factors (input sizes, selectivity, output width,
parallelism), and \emph{simple operators} (scan, filter, eval-scalar, sort),
whose CPU time is largely determined by input cardinality and operator type.

\subsubsection{\textbf{Join CPU Time Prediction}}
To predict join CPU time, we derive features that correspond to the matching and
materialization phases in \Cref{sec:join_model}. Typical features include:
build/probe cardinalities ($|R_B|, |R_P|$), their ratio ($|R_B|/|R_P|$) as a proxy
for selectivity, $\log |R_B|$ for non-linear effects, and width-aware terms such
as $|R_{\text{out}}|\cdot \bar{w}_{\text{out}}$ to approximate materialization
cost.

Rather than fixing a single parametric form, we train a Gradient Boosted
Decision Tree (GBDT) regressor~\cite{GBM} to capture non-linear interactions and
threshold effects commonly observed in join execution (e.g., skew, abrupt cache
effects, and memory locality changes).
Training data are collected by executing a diverse set of profiling queries in
$\mathbf{D}$ and extracting operator-level runtime statistics from the engine
profiles; the target variable is the measured CPU time of each join operator.

\stitle{Obtaining cardinalities.}
At inference time, join prediction requires cardinality-related features.
These can be obtained from (i) optimizer estimates, (ii) learned cardinality
models~\cite{MSCN,deepdb,dutt2020efficiently,kiefer2017estimating,woltmann2019cardinality,yang2020neurocard},
or (iii) lightweight auxiliary execution.
In this work, we use an explicit engineering trade-off: we issue an auxiliary
\texttt{COUNT(*)} query that shares the same join graph and predicates but avoids
materializing the full result, providing accurate intermediate cardinalities at
modest overhead. This improves prediction quality for CPU-time matching while
keeping the cost far below running full candidates throughout the search.

\subsubsection{\textbf{CPU Time Prediction for Other Operators}}
For non-join operators, we use lightweight parametric models trained from local
profiles. For operators such as \texttt{Filter} and \texttt{EvalScalar}, we
model CPU time as linear in input cardinality:
\begin{equation}
\label{eq:cpu_linear}
CPU_{\text{op}} = \alpha \cdot |R_{\text{in}}| + \beta,
\end{equation}
with coefficients learned from profiling data in $\mathbf{D}$.
For \texttt{Sort}, we use a super-linear model that incorporates the theoretical
$\Theta(n\log n)$ term and an optional width term:
\begin{equation}
\label{eq:cpu_sort2}
CPU_{\text{sort}}
=
\alpha \cdot |R_{\text{in}}| \cdot \log |R_{\text{in}}|
+ \beta \cdot |R_{\text{in}}| \cdot \bar{w}_{\text{out}}
+ \gamma.
\end{equation}

\stitle{{How \sys uses local models.}}
During realistic cost-aware space bounding (\Cref{sec:cost_aware_bounding}),
$\widehat{CPU}(\mathbf{G})$ screens infeasible candidates early and guides CPU
cost compensation decisions. During predicate search
(\Cref{sec:predicate_search}), local predictions provide cheap feedback for BO
exploration, and the system executes queries in $\mathbf{D}$ only for candidates
whose predicted CPU time falls near the target region, thereby reducing the
number of expensive executions while maintaining accuracy near the optimum.
\section{Introduction of Bendset}
\label{sec:bendset}

Existing public traces such as Snowset~\cite{snowset} and Redset~\cite{redset}
have significantly advanced trace-driven workload research. However, they are
not fully sufficient for \emph{fine-grained query generation} as studied in this
paper. In particular, query-level synthesis benefits from (i) execution targets
that are stable across systems (e.g., CPU time and scanned bytes), (ii)
\emph{structural signals} that can be enforced during generation (e.g., operator
composition), and (iii) repetition-related signals to
reproduce the strong locality observed in production workloads. To better
support these needs and enable reproducible research, we publicly release a new
trace, \textsc{Bendset}.

\subsection{\textsc{Bendset} overview.}
\textsc{Bendset} is a publicly available workload trace released by Databend
Cloud~\cite{databend}. It captures real production queries from 100 serverless
instances over an 8-day period. Unlike the general-purpose analytical workloads
in Snowset and Redset, \textsc{Bendset} reflects a gaming analytics pipeline
that continuously ingests data, performs incremental stream-based
transformations, and serves low-latency analytical queries to produce fresh
features for downstream AI applications. The full trace contains 40 million
queries and 30 per-query features, covering query timing, I/O statistics, memory
consumption, and operator-level metrics.


\begin{table}[t]
\centering
\caption{Comparison of Public Cloud Workload Traces}
\vspace{-0.8em}
\label{tab:workload-comparison}
\setlength{\tabcolsep}{3.5pt}
\begin{tabular}{lccc}
\toprule
 & \textbf{\textsc{Bendset}} & \textbf{Redset} & \textbf{Snowset} \\
\midrule
Source           & Databend Cloud   & Amazon Redshift & Snowflake \\
Queries          & 40M              & $\sim$50M       & $\sim$70M \\
Period           & 8 days           & 3 months        & 14 days \\
Features         & 30               & 24              & 91 \\
Repetition       & 85.0\%           & 42.3\%          & -- \\
Operator signals & Counts           & Counts          & Profiled time \\
\bottomrule
\end{tabular}
\vspace{-0.6em}
\end{table}

\subsection{Comparison with Snowset and Redset.}
\Cref{tab:workload-comparison} summarizes the key characteristics of the
three public traces used in our evaluation. \textsc{Bendset} exhibits a
substantially higher repetition rate (85.0\%) than Redset (42.3\%), which makes
it particularly suitable for studying repetition-aware synthesis and replay. Importantly, \textsc{Bendset} also provides both a
\emph{query hash} and a \emph{parameterized query hash}, enabling accurate
measurement and reproduction of workload repetition at the levels of exact
queries and templates, respectively.
In addition, \textsc{Bendset} and Redset provide explicit operator statistics
(e.g., counts of joins/aggregations/sorts), whereas Snowset mainly reports
coarse operator-level profiled time; this difference is important for query
generation because operator \emph{counts} can be directly enforced as
structure-aware constraints.
\section{Experiments}
\label{sec:eval}
We evaluate \sys on a real cloud database service and three public workload
traces. Unless otherwise stated, all reported numbers are measured end-to-end,
including generation, auxiliary executions, and metric collection.
\subsection{Experiment Setup}
\label{sec:exp_setup}
\noindent \textbf{Platform and Implementation.}
    We run all experiments on Databend Cloud~\cite{databend}, a Snowflake-like
    cloud data warehouse service.
    Unless noted otherwise, each run uses a \emph{Small} warehouse configuration.
    We collect query execution metrics from \texttt{system\_history.query\_history}.
    We use \texttt{skopt}~\cite{skopt} as the Bayesian Optimization (BO) solver.
  We use the
    latest \texttt{qwen-plus} model   for query graph translator (~\Cref{sec:LLM_trans}).

    \noindent \textbf{Workload Traces.} 
      We evaluate on three public cloud workload traces: \textsc{Bendset},
    \textsc{Redset}~\cite{redset}, and \textsc{Snowset}~\cite{snowset}.
    All traces contain only \texttt{SELECT} queries, and we preserve the original
    temporal order in each sampled trace to reflect realistic workload dynamics.
    \emph{Bendset:} We randomly sample two temporally consecutive 2-hour traces
    from two different days. Each trace contains 200 queries executed in
    sequence. The two traces exhibit different repetition rates (40.5\% and
    8.5\%, respectively).
    \emph{Redset:} We randomly select one 1-hour trace containing 104 consecutive
    queries; its repetition rate is 42.3\%.
    \emph{Snowset:} We randomly extract one trace consisting of 54 queries
    spanning approximately 30 minutes.

    \noindent \textbf{Initial Workload.}   We instantiate the testing dataset using benchmark databases with
    multiple scale factors.
    Specifically, we deploy TPC-H at 1GB, 2GB, and 5GB, and TPC-DS at 1GB and
    2GB. 
    
    \noindent \textbf{Evaluation Metrics.} \emph{(1) Performance accuracy.} For execution targets in $\mathbf{Y}$ (CPU time
    and scanned bytes), we report \emph{q-error} and summarize results using the
    median, 90th, and 99th percentiles to capture both typical and tail
    behaviors.
    \emph{(2) Structural adherence.} For each structure-aware constraint in
    $\mathbf{C}$ (e.g., join/aggregate/sort counts), we report per-query absolute
    error and the average MAE over the evaluated workload.
    \emph{(3) LLM cost.} We report prompt and completion tokens consumed by
    LLM-based components.
    \emph{(4) Generation latency.} We report end-to-end wall-clock time,
    including BO iterations, local model inference, and auxiliary query
    execution.

\begin{figure*}
    \centering
    \vspace{-2em}
    \includegraphics[width=0.90\linewidth]{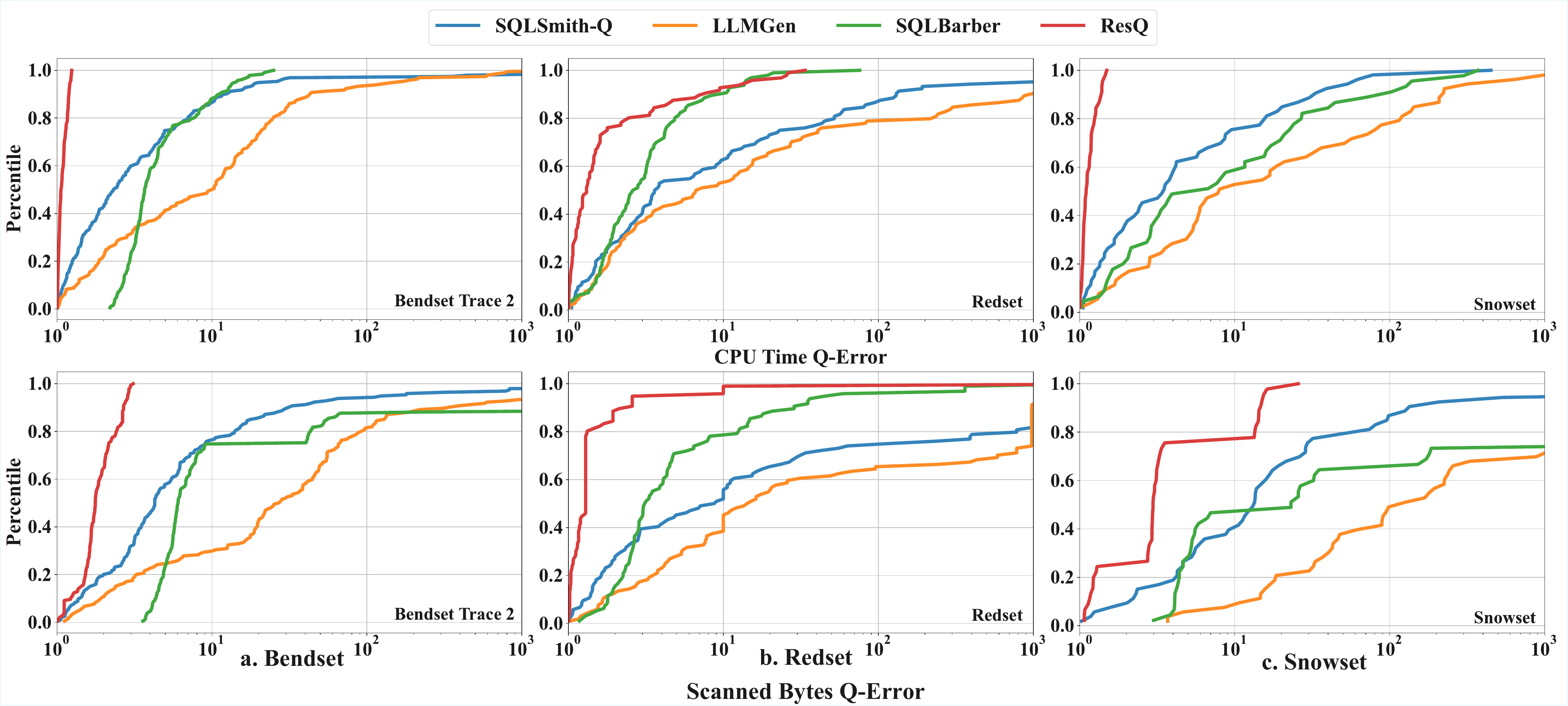}
    \vspace{-1em}
    \caption{Q-error CDFs for \texttt{CPUTime} (top) and \texttt{ScannedBytes} (bottom): (a) {Bendset} Trace~2, (b) {Redset} trace, and (c) {Snowset} trace.}
    \vspace{-1em}
    \label{fig:exp_cdf}
\end{figure*}



    \noindent \textbf{Baseline.} We evaluate Three baselines as follows: 
    \textbf{(1) SQLSmith-Q~\cite{seltenreich2020sqlsmith}:} It generates syntactically valid SQL queries by randomly expanding a parse tree, and is primarily designed for bug finding.
    For query-level evaluation, we adapt it as \emph{SQLSmith-Q}, which randomly generates three queries for each target and constraint setting to assess how well random generation can match execution objectives.
    \textbf{(2) LLMGen:} LLMGen is adapted from the LLM-based augmentation module in PBench~\cite{pbench}, which employs multi-round conversational rewriting guided by expert knowledge to generate queries that approximate given execution targets and structural constraints.
    Following PBench, we use three iterative refinement rounds for each input target.
    \textbf{(3) SQLBarber \cite{sqlbarber}:} SQLBarber~\cite{sqlbarber} synthesizes workloads by partitioning a workload’s logical cost distribution into disjoint cost ranges, generating SQL templates using an LLM, and applying Bayesian Optimization (BO) to search predicates that allow each template to cover its assigned range.
    To align SQLBarber with query-level generation, we adapt it to generate one template per target query and apply BO-based predicate search to instantiate the template for each execution target. 


\subsection{Evaluation On Bendset}

\begin{table*}[h]
\caption{Evaluation of \sys on Bendset trace 1.}
 \vspace{-1em}
\label{tab:bendset1}
\centering
\setlength{\tabcolsep}{3pt}
\begin{tabular}{|c|ccc|ccc|c|c|c|c|c|c|}
\hline
\multirow{2}{*}{Methods} & \multicolumn{3}{c|}{CPU Time Q-Error}                                                   & \multicolumn{3}{c|}{Scanned Bytes Q-Error}                                                      & \multicolumn{1}{l|}{\multirow{2}{*}{\makecell{Join\\ MAE}}} & \multicolumn{1}{l|}{\multirow{2}{*}{\makecell{Aggregate\\MAE}}} & \multicolumn{1}{l|}{\multirow{2}{*}{\makecell{Sort\\ MAE}}} & \multirow{2}{*}{\makecell{Prompt Tokens\\(K)}} & \multirow{2}{*}{\makecell{Output Tokens\\(K)}} & \multirow{2}{*}{\makecell{Latency\\(min)}} \\ \cline{2-7}
                         & \multicolumn{1}{c|}{Median}        & \multicolumn{1}{c|}{90th}          & 99th          & \multicolumn{1}{c|}{Median}        & \multicolumn{1}{c|}{90th}          & 99th          & \multicolumn{1}{l|}{}                          & \multicolumn{1}{l|}{}                               & \multicolumn{1}{l|}{}                          &                                    &                                    &                                \\ \hline
SQLSmith-Q                 & \multicolumn{1}{c|}{2.19}          & \multicolumn{1}{c|}{58.52}         & 425.59        & \multicolumn{1}{c|}{3.74}          & \multicolumn{1}{c|}{137.51}        & 9247.89       & \textbf{0.00}                                  & 0.55                                                & \textbf{0.00}                                  & \textbf{0.00}& \textbf{0.00}& \textbf{202.07}                \\ \hline
LLMGen                   & \multicolumn{1}{c|}{5.53}          & \multicolumn{1}{c|}{8.69}          & 19.73         & \multicolumn{1}{c|}{129.74}        & \multicolumn{1}{c|}{5367.18}       & 8550.57       & 0.03                                           & 0.90                                                & 0.47& 2244.19                            & 265.52                             & 249.25                         \\ \hline
SQLBarber                & \multicolumn{1}{c|}{2.31}          & \multicolumn{1}{c|}{4.29}          & 11.029        & \multicolumn{1}{c|}{2.41}          & \multicolumn{1}{c|}{4.08}          & 9.93          & 0.07                                           & \textbf{0.02}                                       & 0.12                                           & 870.51                             & 510.67                             & 1560.85                        \\ \hline
\sys                     & \multicolumn{1}{c|}{\textbf{1.13}} & \multicolumn{1}{c|}{\textbf{1.42}} & \textbf{1.66} & \multicolumn{1}{c|}{\textbf{1.14}} & \multicolumn{1}{c|}{\textbf{1.69}} & \textbf{2.56} & \textbf{0.00}                                  & 0.07                                                & \textbf{0.00}                                  & 144.32                             & 2.83                               & 281.26                         \\ \hline
\end{tabular}
\end{table*}

\begin{table*}[h]
\caption{Evaluation of \sys on Bendset trace 2.}
 \vspace{-1em}
\label{tab:bendset2}
\centering
\setlength{\tabcolsep}{3pt}
\begin{tabular}{|c|ccc|ccc|c|c|c|c|c|c|}
\hline
\multirow{2}{*}{Methods} & \multicolumn{3}{c|}{CPU Time Q-Error}                                                   & \multicolumn{3}{c|}{Scanned Bytes Q-Error}                                                      & \multicolumn{1}{l|}{\multirow{2}{*}{\makecell{Join\\ MAE}}} & \multicolumn{1}{l|}{\multirow{2}{*}{\makecell{Aggregate\\MAE}}} & \multicolumn{1}{l|}{\multirow{2}{*}{\makecell{Sort\\ MAE}}} & \multirow{2}{*}{\makecell{Prompt Tokens\\(K)}} & \multirow{2}{*}{\makecell{Output Tokens\\(K)}} & \multirow{2}{*}{\makecell{Latency\\(min)}} \\ \cline{2-7}
                         & \multicolumn{1}{c|}{Median}        & \multicolumn{1}{c|}{90th}          & 99th          & \multicolumn{1}{c|}{Median}        & \multicolumn{1}{c|}{90th}          & 99th          &                           &                                &                           &                                    &                                    &                                \\ \hline
SQLSmith-Q                 & \multicolumn{1}{c|}{2.30}          & \multicolumn{1}{c|}{11.44}         & 522.56        & \multicolumn{1}{c|}{4.27}          & \multicolumn{1}{c|}{28.95}         & 5319.80       & \textbf{0.00}             & 0.27                           & \textbf{0.00}             & \textbf{0.00}                      & \textbf{0.00}                      & 175.51                         \\ \hline
LLMGen                   & \multicolumn{1}{c|}{8.80}          & \multicolumn{1}{c|}{32.92}         & 148.41        & \multicolumn{1}{c|}{24.17}         & \multicolumn{1}{c|}{133.41}        & 1346.78       & \textbf{0.00}             & 0.71                           & 0.75                      & 4487.67                            & 383.65                             & 313.51                         \\ \hline
SQLBarber                & \multicolumn{1}{c|}{3.66}          & \multicolumn{1}{c|}{10.63}         & 17.92         & \multicolumn{1}{c|}{6.00}          & \multicolumn{1}{c|}{4628.11}       & 5716.56       & \textbf{0.00}             & \textbf{0.00}                  & 0.02                      & 2054.13                             & 1665.9                             & 1454.27                        \\ \hline
\sys                     & \multicolumn{1}{c|}{\textbf{1.05}} & \multicolumn{1}{c|}{\textbf{1.18}} & \textbf{1.21} & \multicolumn{1}{c|}{\textbf{1.76}} & \multicolumn{1}{c|}{\textbf{2.64}} & \textbf{2.87} & \textbf{0.00}             & 0.08                           & \textbf{0.00}             & 18.97                              & 2.28                               & \textbf{111.73}                \\ \hline
\end{tabular}

\end{table*}

We first evaluate \sys on two \textsc{Bendset} traces sampled as described in
\Cref{sec:exp_setup}. We report per-query target accuracy, adherence to
structure-aware constraints (MAE over operator counts), LLM token consumption,
and end-to-end generation latency. \Cref{fig:exp_cdf} further summarizes the
error distributions using CDF plots.

\stitle{Performance Accuracy.} \Cref{tab:bendset1,tab:bendset2} show that \sys consistently achieves the best
target-fitting accuracy across both CPU time and scanned bytes.
In Trace 1, our method achieves a median CPU Time Q-error of 1.13, outperforming SQLBarber (2.31) and SQLSmith-Q (2.19) by a factor of nearly $2\times$. 
The superiority becomes even more pronounced when considering the \textit{Scanned Bytes} metric in Trace 2, where \sys yields a median Q-error of 1.76, while SQLBarber and LLMGen struggle with errors of 6.00 and 24.17, respectively. 
These results indicate that our approach possesses a fine-grained understanding of the relationship between query structure and physical resource consumption, whereas baselines often produce queries that deviate significantly from the intended performance profile. 

\stitle{Robustness at the Tail.}
\sys also exhibits significantly better tail behavior.
On Trace~2, SQLSmith-Q’s 99th-percentile CPU-time q-error reaches 522.56, whereas
\sys remains at 1.21. 
Similarly, for Scanned Bytes in Trace 1, the 99th percentile error of SQLSmith-Q (9247.89) and SQLBarber (9.93) shows a massive deviation, whereas \sys remains stable at 2.56. 
This near-flat error curve from Median to 99th percentile highlights the robustness and predictability of our generation method. Unlike LLM-based approaches (e.g., LLMGen, SQLBarber), which, due to their intrinsic black-box nature, lack a principled understanding of the underlying physical execution engine and data distribution. Consequently, they rely on a coarse-grained approximation of query performance, which leads to poor robustness when fitting fine-grained real-world targets.

\stitle{Structural Constraint Adherence} To verify whether the generated queries maintain logical consistency while fitting performance targets, we report the Mean Absolute Error (MAE) for key operators. As shown in Tables~\ref{tab:bendset1} and ~\ref{tab:bendset2}, \sys achieves near-zero MAE for Join, Aggregate, and Sort operators (e.g., 0.00 for Join and Sort in both traces). These results demonstrate that our approach successfully adheres to predefined structural constraints, ensuring that the synthesized queries are not only performance-aligned but also structurally consistent with the task specifications.

\stitle{LLM Cost.}
\sys uses the LLM primarily for query-graph-to-SQL translation, resulting in
substantially lower token consumption than LLM-based baselines. On Trace~2, \sys consumes 18.97K prompt tokens, representing a reduction of over two orders of magnitude relative to LLMGen (4487.67K) and SQLBarber (2054.13K). This disparity arises because baselines must ingest extensive few-shot examples, intricate table schemas, and complex data distribution statistics within the prompt to guide the generation process. In contrast, \sys simplifies the LLM's role to a focused translation task—mapping pre-optimized operator trees to SQL syntax—thereby drastically shrinking the input context. Furthermore, the reliance of SQLBarber and LLMGen on iterative multi-turn dialogues for query rewriting and error correction leads to significantly inflated completion tokens (e.g., 1665.9K for SQLBarber vs. our 2.28K), whereas \sys achieves high-fidelity synthesis in a single pass. This "single-pass translation" paradigm not only eliminates the need for costly refinement cycles but also ensures superior scalability for large-scale query generation tasks where API costs and latency are critical constraints.

\stitle{Efficiency.}
To evaluate the practical scalability of \sys, we analyze the end-to-end latency alongside resource consumption. Table~\ref{tab:bendset1} and \ref{tab:bendset1} report the total wall-clock time required for query generation across different traces. Our method demonstrates significant efficiency gains, particularly when compared to baselines that involve iterative refinement(e.g., BO predicate search in SQLBarber). In Trace 2, \sys completes the generation task in only 111.73 minutes, achieving a 13$\times$ speedup over SQLBarber (1454.27 minutes). While simpler methods like SQLSmith-Q may exhibit comparable latency in some scenarios due to their template-based nature, they fail to meet the performance targets. Notably, \sys still delivers a 5.5$\times$ speedup over SQLBarber in Trace 1 while maintaining superior optimization precision. The substantial latency gap stems from the fundamental difference in generation paradigms; specifically, SQLBarber lacks a search space bounding mechanism and incurs significant overhead by utilizing actual execution feedback during its BO process.

Finally, \sys benefits further from repetition.
On Trace~2 (40.5\% repetition), query-pool reuse amortizes the cost of structure
construction and predicate search across repeated queries/templates, improving
throughput and making the advantage more pronounced for longer traces.

\subsection{Evaluation On Redset}
To further evaluate the generalization of \sys, we report its performance on the Redset trace in ~\Cref{fig:exp_cdf} and ~\Cref{tab:redset}. The results are highly consistent with the observations in previous traces, further validating the robustness of our approach.

\stitle{Target Accuracy and Tail Robustness.} As shown in Table~\ref{tab:redset}, \sys continues to demonstrate superior target-fitting accuracy. Notably, for \textit{Scanned Bytes}, our method achieves a 99th percentile Q-error of only 1.93, standing in sharp contrast to SQLSmith-Q (4666.01) and LLMGen (985.87). Even in the more volatile \textit{CPU Time} metric, \sys maintains a tight 99th percentile bound of 12.61. This confirms that our framework reliably handles diverse data distributions and complex query patterns. In terms of generation efficiency, \sys completes the task in 126.31 minutes, achieving a 6.1$\times$ speedup over SQLBarber.

\stitle{Structural Constraint Adherence.}
\sys maintains strict adherence to structure-aware constraints, achieving a Join
MAE of 0.00 and an Aggregate MAE of 0.02 in \Cref{tab:redset}. This indicates
that improved target matching does not come at the cost of violating the
trace-derived structural profile.

\stitle{Efficiency.}
\sys completes the \textsc{Redset} trace in 126.31 minutes, achieving a
$6.1\times$ speedup over SQLBarber (773.49 minutes) while delivering better
target accuracy. This efficiency stems from \sys’s bounded search and reduced
reliance on expensive query executions during BO.


\begin{table*}[h]
\caption{Evaluation of \sys on Redset trace.}
 \vspace{-1em}
\label{tab:redset}
\centering
\setlength{\tabcolsep}{3pt}
\begin{tabular}{|c|ccc|ccc|c|c|c|c|c|}
\hline
\multirow{2}{*}{Methods} & \multicolumn{3}{c|}{CPU Time Q-Error}                                                    & \multicolumn{3}{c|}{Scanned Bytes Q-Error}                                                      & \multicolumn{1}{l|}{\multirow{2}{*}{\makecell{Join\\MAE}}} & \multicolumn{1}{l|}{\multirow{2}{*}{\makecell{Aggregate\\MAE}}} & \multirow{2}{*}{\makecell{Prompt Tokens\\(K)}} & \multirow{2}{*}{\makecell{Output Tokens\\(K)}} & \multirow{2}{*}{\makecell{Latency\\(min)}} \\ \cline{2-7}
                         & \multicolumn{1}{c|}{Median}        & \multicolumn{1}{c|}{90th}          & 99th           & \multicolumn{1}{c|}{Median}        & \multicolumn{1}{c|}{90th}          & 99th          & \multicolumn{1}{l|}{}                          & \multicolumn{1}{l|}{}                               &                                    &                                    &                                \\ \hline
SQLSmith-Q                 & \multicolumn{1}{c|}{3.59}          & \multicolumn{1}{c|}{59.11}         & 1599.00        & \multicolumn{1}{c|}{7.90}          & \multicolumn{1}{c|}{976.56}        & 4666.01       & 0.00                                           & 0.43                                                & \textbf{0.00}                      & \textbf{0.00}                      & 132.42                \\ \hline
LLMGen                   & \multicolumn{1}{c|}{5.89}          & \multicolumn{1}{c|}{279.41}        & 1068.62        & \multicolumn{1}{c|}{10.97}         & \multicolumn{1}{c|}{976.56}        & 985.87        & 0.03                                           & 0.19                                                & 1026.92                            & 84.13                              & {128.85}                \\ \hline
SQLBarber                & \multicolumn{1}{c|}{2.45}          & \multicolumn{1}{c|}{5.82}          & 13.59          & \multicolumn{1}{c|}{3.71}          & \multicolumn{1}{c|}{16.17}         & 36.23         & 0.14                                           & 0.04& 768.36                             & 72.52                              & 773.49                         \\ \hline
\sys                     & \multicolumn{1}{c|}{\textbf{1.23}} & \multicolumn{1}{c|}{\textbf{4.35}} & \textbf{12.61} & \multicolumn{1}{c|}{\textbf{1.18}} & \multicolumn{1}{c|}{\textbf{1.30}} & \textbf{1.93} & \textbf{0.00}                                  & \textbf{0.02}& 123.86                    & 2.97                      & \textbf{126.31  }                       \\ \hline
\end{tabular}
\end{table*}

\subsection{Evaluation On Snowset}
Finally, we evaluate \sys on Snowset. Compared with \textsc{Bendset} and
\textsc{Redset}, \textsc{Snowset} provides coarser structural signals (operator
profiled time rather than explicit operator counts), making fine-grained query
generation more challenging.

\stitle{Target Accuracy and Stability.}
As shown in \Cref{tab:snowset}, baselines become substantially less stable under
this reduced structural information, especially for scanned bytes: the
99th-percentile q-error reaches 7769.17 for LLMGen and 4807.46 for SQLBarber.
In contrast, \sys remains stable, achieving a 99th-percentile CPU-time q-error of
1.25 and a 99th-percentile scanned-bytes q-error of 14.24. These results
highlight the benefit of \sys’s search-space bounding, which constrains the
optimization space using data-access feasibility rather than relying solely on
precise operator-count constraints.

\stitle{Reuse without explicit hashes.}
Unlike \textsc{Bendset}, \textsc{Snowset} does not provide query hash or
parameterized query hash fields. To still benefit from reuse, \sys performs
retrieval using the available target/constraint signals (e.g., operator
presence indicators and execution-target ranges) to identify previously
generated queries/templates with similar profiles, and reuses them as warm
starts for generation. This reuse contributes to both stability and efficiency
even when exact repetition identifiers are unavailable.

\stitle{Efficiency.}
\sys achieves the lowest end-to-end latency on \textsc{Snowset} (53.47 minutes),
outperforming SQLBarber (453.86 minutes) and remaining competitive with the
lighter baselines while providing much higher target accuracy.


\begin{table*}[h]
\caption{Evaluation of \sys on Snowset trace.}
 \vspace{-1em}
\label{tab:snowset}
\centering
\setlength{\tabcolsep}{3pt}
\begin{tabular}{|c|ccc|ccc|c|c|c|c|c|c|}
\hline
\multirow{2}{*}{Methods} & \multicolumn{3}{c|}{CPU Time Q-Error}                                                   & \multicolumn{3}{c|}{Scanned Bytes Q-Error}                                                       & \multicolumn{1}{l|}{\multirow{2}{*}{\makecell{Join\\ MAE}}} & \multicolumn{1}{l|}{\multirow{2}{*}{\makecell{Aggregate\\MAE}}} & \multicolumn{1}{l|}{\multirow{2}{*}{\makecell{Sort\\ MAE}}} & \multirow{2}{*}{\makecell{Prompt Tokens\\(K)}} & \multirow{2}{*}{\makecell{Output Tokens\\(K)}} & \multirow{2}{*}{\makecell{Latency\\(min)}} \\ \cline{2-7}
                         & \multicolumn{1}{c|}{Median}        & \multicolumn{1}{c|}{90th}          & 99th          & \multicolumn{1}{c|}{Median}        & \multicolumn{1}{c|}{90th}          & 99th           & \multicolumn{1}{l|}{}                          & \multicolumn{1}{l|}{}                               & \multicolumn{1}{l|}{}                          &                                    &                                    &                                \\ \hline
SQLSmith-Q                 & \multicolumn{1}{c|}{2.42}          & \multicolumn{1}{c|}{25.48}         & 57.63& \multicolumn{1}{c|}{13.19}         & \multicolumn{1}{c|}{98.27}         & 24078.12& \textbf{0.00}                                  & 0.46& \textbf{0.00}                                  & \textbf{0.00}                      & \textbf{0.00}                      & 60.58                          \\ \hline
LLMGen                   & \multicolumn{1}{c|}{5.97}          & \multicolumn{1}{c|}{84.88}         & 140.27        & \multicolumn{1}{c|}{89.19}         & \multicolumn{1}{c|}{3865.10}       & 7769.17        & 0.11                                           & 0.07                                                & \textbf{0.00}                                  & 894.79                             & 79.76                              & 63.92                          \\ \hline
SQLBarber                & \multicolumn{1}{c|}{3.12}          & \multicolumn{1}{c|}{17.64}         & 27.66         & \multicolumn{1}{c|}{25.72}         & \multicolumn{1}{c|}{172.33}        & 4807.46        & 0.21                                           & 0.35                                                & \textbf{0.00}                                  & 183.10                             & 127.40                             & 453.86                         \\ \hline
\sys                     & \multicolumn{1}{c|}{\textbf{1.06}} & \multicolumn{1}{c|}{\textbf{1.18}} & \textbf{1.25} & \multicolumn{1}{c|}{\textbf{2.95}} & \multicolumn{1}{c|}{\textbf{9.52}} & \textbf{14.24} & \textbf{0.00}                                  & \textbf{0.05}& \textbf{0.00}                                  & 9.17                               & 0.87                               & \textbf{53.47}                 \\ \hline
\end{tabular}
\end{table*}

\subsection{Ablation Study}
\label{sec:ablation}
\begin{figure}[h]
    \centering
    \vspace{-1em}
    \includegraphics[width=0.95\linewidth]{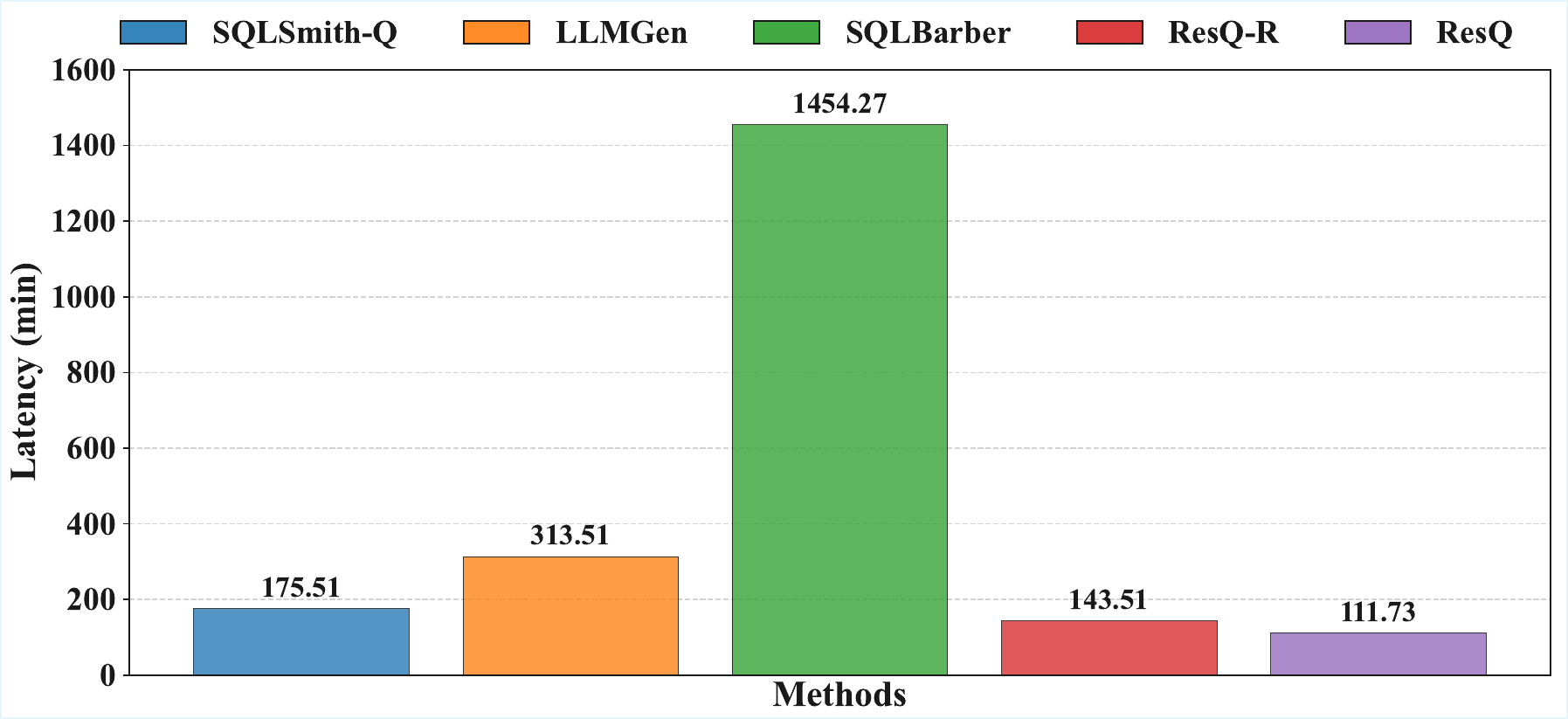}
    \vspace{-1em}
    \caption{Ablation Study of Local Model}
    \vspace{-2em}
    \label{fig:ablation}
\end{figure}

\stitle{Impact of local performance models.}
As shown in Figure~\ref{fig:ablation}, we evaluate the local model's efficiency on Bendset trace 2, the inclusion of the local model significantly enhances the generation efficiency. Specifically, the end-to-end latency of \sys is 111.73 minutes, whereas the ablated version \sys-R requires 143.51 minutes to complete the same task. Which indicates that the local model significantly reduces the times of real execution during the BO process. 

The efficiency gain stems from the local model's ability to act as a high-speed proxy during the Bayesian Optimization phase. By providing rapid performance estimates for candidate query structures, the local model allows the system to prune unpromising regions of the search space without invoking more expensive evaluation steps. Consequently, \sys achieves a much higher throughput than \sys-R and consistently outperforms all baseline methods, including SQLSmith-Q (175.51 min), LLMGen (313.51 min), and SQLBarber (1454.27 min), in terms of time-to-delivery. This underscores that the local model is a critical component for making target-driven query generation practical for large-scale database testing environments.

\section{Related Work}
\label{sec:related}

Existing fine-grained query generation methods can be broadly classified into two lines.  

\vspace{0.5mm}\noindent
\textbf{(1) Optimizer-based cost-aware query generation.}
These methods generate SQL queries to match a \emph{single} Optimizer-based cost target, such as
result cardinality or plan cost, with the goal of producing diverse test
queries that stress a DBMS.
They typically fall into three categories.
(i) \emph{Random generation.}
SQLSmith~\cite{seltenreich2020sqlsmith} is a classic random query generator that quickly produces large numbers of queries for a given database, but often yields invalid queries and provides limited controllability for meeting specific targets.
Bati et al.~\cite{Bati_Giakoumakis_Herbert_Surna_2007} use genetic algorithms to improve coverage and efficiency of random test generation, but do not explicitly target key cost constraints (e.g., cardinality) or structural requirements.
(ii) \emph{Template-based generation.}
This line first constructs query templates subject to structural constraints, and then tunes predicates to satisfy a target cardinality/cost range.
Bruno et al.~\cite{bruno} show that logical constraint-aware query generation is NP-hard and propose hill-climbing over predicate values for cardinality targeting.
Mishra et al.~\cite{Mishra_Koudas_Zuzarte_2008} extend the setting to multiple cardinality targets and introduce pruning algorithm to reduce optimization cost.
More recently, SQLBarber~\cite{sqlbarber} uses an LLM to iteratively propose and refine SQL templates under user-specified structural requirements (e.g., operator counts), and then applies Bayesian optimization to tune predicates toward target cost/cardinality ranges.
(iii) \emph{Learning-based generation.}
LearnedSQLGen~\cite{Learnedsqlgen} trains a reinforcement learning agent to generate SQL token-by-token, using logical cost estimates as rewards and a finite-state machine to enforce syntactic validity.

\vspace{0.5mm}\noindent
\textbf{(2) Realistic performance-aware query generation.}
This line targets \emph{execution-level} signals observed in cloud production workloads, such as CPU time, scanned bytes, memory/I/O usage, and physical operator distributions.
Because production SQL and user data are typically unavailable for privacy reasons, these methods aim to synthesize executable queries that reproduce similar execution behaviors in a controlled environment based on anonymized performance traces.
Zhou et al.~\cite{pbench} propose an LLM-based, multi-target SQL generation workflow in PBench that iteratively refines candidates using multi-round interactions and feedback from query execution.
However, in PBench this generation component is mainly used to expand a query pool for coarse-level workload synthesis, and it does not explicitly optimize for or systematically evaluate per-query execution-profile matching accuracy.

\vspace{0.5mm}\noindent
\textbf{Positioning.}
The realistic performance-aware setting differs fundamentally from targeted query generation~\cite{bruno,Mishra_Koudas_Zuzarte_2008,sqlbarber,Learnedsqlgen}:
instead of matching a single logical plan/optimizer-level cost target, our goal is to match \emph{multiple} observed execution-level metrics per query (and their repetition patterns) under privacy and proxy-database constraints.

\section{Conclusion}

\label{sec:con}
We presented \sys for fine-grained, trace-driven workload synthesis under
privacy constraints, generating executable SQL queries that match per-query
execution targets and structural signals. By combining two-phase query-graph
generation, performance-aware bounding, lightweight local models, and
repetition-aware reuse, \sys achieves high fidelity and efficiency on Snowset,
Redset, and the newly released \textsc{Bendset}.

%

\clearpage

\balance

\bibliographystyle{ACM-Reference-Format}
\bibliography{references}


\begin{thebibliography}{31}


\ifx \showCODEN    \undefined \def \showCODEN     #1{\unskip}     \fi
\ifx \showDOI      \undefined \def \showDOI       #1{#1}\fi
\ifx \showISBNx    \undefined \def \showISBNx     #1{\unskip}     \fi
\ifx \showISBNxiii \undefined \def \showISBNxiii  #1{\unskip}     \fi
\ifx \showISSN     \undefined \def \showISSN      #1{\unskip}     \fi
\ifx \showLCCN     \undefined \def \showLCCN      #1{\unskip}     \fi
\ifx \shownote     \undefined \def \shownote      #1{#1}          \fi
\ifx \showarticletitle \undefined \def \showarticletitle #1{#1}   \fi
\ifx \showURL      \undefined \def \showURL       {\relax}        \fi
\providecommand\bibfield[2]{#2}
\providecommand\bibinfo[2]{#2}
\providecommand\natexlab[1]{#1}
\providecommand\showeprint[2][]{arXiv:#2}

\bibitem[\protect\citeauthoryear{Bati, Giakoumakis, Herbert, and Surna}{Bati et~al\mbox{.}}{2007}]%
        {Bati_Giakoumakis_Herbert_Surna_2007}
\bibfield{author}{\bibinfo{person}{Hardik Bati}, \bibinfo{person}{Leo Giakoumakis}, \bibinfo{person}{Steve Herbert}, {and} \bibinfo{person}{Aleksandras Surna}.} \bibinfo{year}{2007}\natexlab{}.
\newblock \showarticletitle{A genetic approach for random testing of database systems}.
\newblock \bibinfo{journal}{\emph{Very Large Data Bases,Very Large Data Bases}} (\bibinfo{date}{Sep} \bibinfo{year}{2007}).
\newblock


\bibitem[\protect\citeauthoryear{Bruno, Chaudhuri, and Thomas}{Bruno et~al\mbox{.}}{2006}]%
        {bruno}
\bibfield{author}{\bibinfo{person}{Nicolas Bruno}, \bibinfo{person}{Surajit Chaudhuri}, {and} \bibinfo{person}{Dilys Thomas}.} \bibinfo{year}{2006}\natexlab{}.
\newblock \showarticletitle{Generating queries with cardinality constraints for dbms testing}.
\newblock \bibinfo{journal}{\emph{IEEE Transactions on Knowledge and Data Engineering}} \bibinfo{volume}{18}, \bibinfo{number}{12} (\bibinfo{year}{2006}), \bibinfo{pages}{1721--1725}.
\newblock


\bibitem[\protect\citeauthoryear{databendlabs}{databendlabs}{2025}]%
        {databend}
\bibfield{author}{\bibinfo{person}{databendlabs}.} \bibinfo{year}{2025}\natexlab{}.
\newblock \bibinfo{title}{Databend: One Rust Warehouse for Analytics, Search, AI}.
\newblock \bibinfo{howpublished}{\url{https://github.com/databendlabs/databend}}.
\newblock
\newblock
\shownote{GitHub repository.}


\bibitem[\protect\citeauthoryear{Ding, Marcus, Kipf, Nathan, Nrusimha, Vaidya, van Renen, and Kraska}{Ding et~al\mbox{.}}{2022}]%
        {ding2022sagedb}
\bibfield{author}{\bibinfo{person}{Jialin Ding}, \bibinfo{person}{Ryan Marcus}, \bibinfo{person}{Andreas Kipf}, \bibinfo{person}{Vikram Nathan}, \bibinfo{person}{Aniruddha Nrusimha}, \bibinfo{person}{Kapil Vaidya}, \bibinfo{person}{Alexander van Renen}, {and} \bibinfo{person}{Tim Kraska}.} \bibinfo{year}{2022}\natexlab{}.
\newblock \showarticletitle{Sagedb: An instance-optimized data analytics system}.
\newblock \bibinfo{journal}{\emph{Proceedings of the VLDB Endowment}} \bibinfo{volume}{15}, \bibinfo{number}{13} (\bibinfo{year}{2022}).
\newblock


\bibitem[\protect\citeauthoryear{Dong, Zhang, Li, and Zhang}{Dong et~al\mbox{.}}{2024}]%
        {chao2024cloud}
\bibfield{author}{\bibinfo{person}{Haowen Dong}, \bibinfo{person}{Chao Zhang}, \bibinfo{person}{Guoliang Li}, {and} \bibinfo{person}{Huanchen Zhang}.} \bibinfo{year}{2024}\natexlab{}.
\newblock \showarticletitle{Cloud-native databases: A survey}.
\newblock \bibinfo{journal}{\emph{IEEE Transactions on Knowledge and Data Engineering}} \bibinfo{volume}{36}, \bibinfo{number}{12} (\bibinfo{year}{2024}), \bibinfo{pages}{7772--7791}.
\newblock


\bibitem[\protect\citeauthoryear{Dutt, Wang, Narasayya, and Chaudhuri}{Dutt et~al\mbox{.}}{2020}]%
        {dutt2020efficiently}
\bibfield{author}{\bibinfo{person}{Anshuman Dutt}, \bibinfo{person}{Chi Wang}, \bibinfo{person}{Vivek Narasayya}, {and} \bibinfo{person}{Surajit Chaudhuri}.} \bibinfo{year}{2020}\natexlab{}.
\newblock \showarticletitle{Efficiently approximating selectivity functions using low overhead regression models}.
\newblock \bibinfo{journal}{\emph{Proceedings of the VLDB Endowment}} \bibinfo{volume}{13}, \bibinfo{number}{12} (\bibinfo{year}{2020}), \bibinfo{pages}{2215--2228}.
\newblock


\bibitem[\protect\citeauthoryear{Fan, Fu, Wang, Zhang, Zuo, Wu, Zhang, Yuan, Ni, Huo, et~al\mbox{.}}{Fan et~al\mbox{.}}{2024}]%
        {fan2024towards}
\bibfield{author}{\bibinfo{person}{Hua Fan}, \bibinfo{person}{Dachao Fu}, \bibinfo{person}{Xu Wang}, \bibinfo{person}{Jiachi Zhang}, \bibinfo{person}{Chaoji Zuo}, \bibinfo{person}{Zhengyi Wu}, \bibinfo{person}{Miao Zhang}, \bibinfo{person}{Kang Yuan}, \bibinfo{person}{Xizi Ni}, \bibinfo{person}{Guocheng Huo}, {et~al\mbox{.}}} \bibinfo{year}{2024}\natexlab{}.
\newblock \showarticletitle{Towards millions of database transmission services in the cloud}.
\newblock \bibinfo{journal}{\emph{Proceedings of the VLDB Endowment}} \bibinfo{volume}{17}, \bibinfo{number}{12} (\bibinfo{year}{2024}), \bibinfo{pages}{4001--4013}.
\newblock


\bibitem[\protect\citeauthoryear{Hilprecht, Schmidt, Kulessa, Molina, Kersting, and Binnig}{Hilprecht et~al\mbox{.}}{2019}]%
        {deepdb}
\bibfield{author}{\bibinfo{person}{Benjamin Hilprecht}, \bibinfo{person}{Andreas Schmidt}, \bibinfo{person}{Moritz Kulessa}, \bibinfo{person}{Alejandro Molina}, \bibinfo{person}{Kristian Kersting}, {and} \bibinfo{person}{Carsten Binnig}.} \bibinfo{year}{2019}\natexlab{}.
\newblock \showarticletitle{DeepDB: Learn from Data, not from Queries!}
\newblock \bibinfo{journal}{\emph{CoRR}}  \bibinfo{volume}{abs/1909.00607} (\bibinfo{year}{2019}).
\newblock
\showeprint[arXiv]{1909.00607}
\urldef\tempurl%
\url{http://arxiv.org/abs/1909.00607}
\showURL{%
\tempurl}


\bibitem[\protect\citeauthoryear{{holgern}}{{holgern}}{2026}]%
        {skopt}
\bibfield{author}{\bibinfo{person}{{holgern}}.} \bibinfo{year}{2026}\natexlab{}.
\newblock \bibinfo{title}{{scikit-optimize}: Sequential model-based optimization with a scipy.optimize interface}.
\newblock \bibinfo{howpublished}{\url{https://github.com/holgern/scikit-optimize}}.
\newblock


\bibitem[\protect\citeauthoryear{Ke, Meng, Finley, Wang, Chen, Ma, Ye, and Liu}{Ke et~al\mbox{.}}{2017}]%
        {GBM}
\bibfield{author}{\bibinfo{person}{Guolin Ke}, \bibinfo{person}{Qi Meng}, \bibinfo{person}{Thomas Finley}, \bibinfo{person}{Taifeng Wang}, \bibinfo{person}{Wei Chen}, \bibinfo{person}{Weidong Ma}, \bibinfo{person}{Qiwei Ye}, {and} \bibinfo{person}{Tie-Yan Liu}.} \bibinfo{year}{2017}\natexlab{}.
\newblock \showarticletitle{Lightgbm: A highly efficient gradient boosting decision tree}.
\newblock \bibinfo{journal}{\emph{Advances in neural information processing systems}}  \bibinfo{volume}{30} (\bibinfo{year}{2017}).
\newblock


\bibitem[\protect\citeauthoryear{Kiefer, Heimel, Bre{\ss}, and Markl}{Kiefer et~al\mbox{.}}{2017}]%
        {kiefer2017estimating}
\bibfield{author}{\bibinfo{person}{Martin Kiefer}, \bibinfo{person}{Max Heimel}, \bibinfo{person}{Sebastian Bre{\ss}}, {and} \bibinfo{person}{Volker Markl}.} \bibinfo{year}{2017}\natexlab{}.
\newblock \showarticletitle{Estimating join selectivities using bandwidth-optimized kernel density models}.
\newblock \bibinfo{journal}{\emph{Proceedings of the VLDB Endowment}} \bibinfo{volume}{10}, \bibinfo{number}{13} (\bibinfo{year}{2017}), \bibinfo{pages}{2085--2096}.
\newblock


\bibitem[\protect\citeauthoryear{Kipf, Kipf, Radke, Leis, Boncz, and Kemper}{Kipf et~al\mbox{.}}{2018}]%
        {MSCN}
\bibfield{author}{\bibinfo{person}{Andreas Kipf}, \bibinfo{person}{Thomas Kipf}, \bibinfo{person}{Bernhard Radke}, \bibinfo{person}{Viktor Leis}, \bibinfo{person}{Peter Boncz}, {and} \bibinfo{person}{Alfons Kemper}.} \bibinfo{year}{2018}\natexlab{}.
\newblock \showarticletitle{Learned cardinalities: Estimating correlated joins with deep learning}.
\newblock \bibinfo{journal}{\emph{arXiv preprint arXiv:1809.00677}} (\bibinfo{year}{2018}).
\newblock


\bibitem[\protect\citeauthoryear{Krid, Stoian, and Kipf}{Krid et~al\mbox{.}}{2025}]%
        {krid2025redbench}
\bibfield{author}{\bibinfo{person}{Skander Krid}, \bibinfo{person}{Mihail Stoian}, {and} \bibinfo{person}{Andreas Kipf}.} \bibinfo{year}{2025}\natexlab{}.
\newblock \showarticletitle{Redbench: A Benchmark Reflecting Real Workloads}.
\newblock \bibinfo{journal}{\emph{arXiv preprint arXiv:2506.12488}} (\bibinfo{year}{2025}).
\newblock


\bibitem[\protect\citeauthoryear{Lao and Trummer}{Lao and Trummer}{2025}]%
        {sqlbarber}
\bibfield{author}{\bibinfo{person}{Jiale Lao} {and} \bibinfo{person}{Immanuel Trummer}.} \bibinfo{year}{2025}\natexlab{}.
\newblock \bibinfo{title}{SQLBarber: A System Leveraging Large Language Models to Generate Customized and Realistic SQL Workloads}.
\newblock
\newblock
\showeprint[arxiv]{2507.06192}~[cs.DB]
\urldef\tempurl%
\url{https://arxiv.org/abs/2507.06192}
\showURL{%
\tempurl}


\bibitem[\protect\citeauthoryear{Lao, Wang, Li, Wang, Zhang, Cheng, Chen, Tang, and Wang}{Lao et~al\mbox{.}}{2025}]%
        {lao2025gptuner}
\bibfield{author}{\bibinfo{person}{Jiale Lao}, \bibinfo{person}{Yibo Wang}, \bibinfo{person}{Yufei Li}, \bibinfo{person}{Jianping Wang}, \bibinfo{person}{Yunjia Zhang}, \bibinfo{person}{Zhiyuan Cheng}, \bibinfo{person}{Wanghu Chen}, \bibinfo{person}{Mingjie Tang}, {and} \bibinfo{person}{Jianguo Wang}.} \bibinfo{year}{2025}\natexlab{}.
\newblock \showarticletitle{GPTuner: An LLM-Based Database Tuning System}.
\newblock \bibinfo{journal}{\emph{ACM SIGMOD Record}} \bibinfo{volume}{54}, \bibinfo{number}{1} (\bibinfo{year}{2025}), \bibinfo{pages}{101--110}.
\newblock


\bibitem[\protect\citeauthoryear{Leis and Kuschewski}{Leis and Kuschewski}{2021}]%
        {Leis_Kuschewski_2021}
\bibfield{author}{\bibinfo{person}{Viktor Leis} {and} \bibinfo{person}{Maximilian Kuschewski}.} \bibinfo{year}{2021}\natexlab{}.
\newblock \showarticletitle{Towards cost-optimal query processing in the cloud}.
\newblock \bibinfo{journal}{\emph{Proceedings of the VLDB Endowment}} \bibinfo{volume}{14}, \bibinfo{number}{9} (\bibinfo{date}{May} \bibinfo{year}{2021}), \bibinfo{pages}{1606–1612}.
\newblock
\urldef\tempurl%
\url{https://doi.org/10.14778/3461535.3461549}
\showDOI{\tempurl}


\bibitem[\protect\citeauthoryear{Li}{Li}{2019}]%
        {li2019cloud}
\bibfield{author}{\bibinfo{person}{Feifei Li}.} \bibinfo{year}{2019}\natexlab{}.
\newblock \showarticletitle{Cloud-native database systems at Alibaba: Opportunities and challenges}.
\newblock \bibinfo{journal}{\emph{Proceedings of the VLDB Endowment}} \bibinfo{volume}{12}, \bibinfo{number}{12} (\bibinfo{year}{2019}), \bibinfo{pages}{2263--2272}.
\newblock


\bibitem[\protect\citeauthoryear{Marcus, Tao, Wu, and Zhao}{Marcus et~al\mbox{.}}{[n.d.]}]%
        {marcussurvivorship}
\bibfield{author}{\bibinfo{person}{Ryan Marcus}, \bibinfo{person}{Jeffrey Tao}, \bibinfo{person}{Peizhi Wu}, {and} \bibinfo{person}{Zijie Zhao}.} \bibinfo{year}{[n.d.]}\natexlab{}.
\newblock \showarticletitle{Survivorship Bias in Industrial Database Workloads}.
\newblock  (\bibinfo{year}{[n.\,d.]}).
\newblock


\bibitem[\protect\citeauthoryear{Misegiannis, Ritter, Giceva, and München}{Misegiannis et~al\mbox{.}}{[n.d.]}]%
        {Misegiannis_Ritter_Giceva_München}
\bibfield{author}{\bibinfo{person}{MichailGeorgoulakis Misegiannis}, \bibinfo{person}{Daniel Ritter}, \bibinfo{person}{Jana Giceva}, {and} \bibinfo{person}{TechnischeUniversität München}.} \bibinfo{year}{[n.d.]}\natexlab{}.
\newblock \showarticletitle{CloudGlide: Deconstructing the Landscape of Cloud-Based Analytics}.
\newblock  (\bibinfo{year}{[n.\,d.]}).
\newblock


\bibitem[\protect\citeauthoryear{Mishra, Koudas, and Zuzarte}{Mishra et~al\mbox{.}}{2008}]%
        {Mishra_Koudas_Zuzarte_2008}
\bibfield{author}{\bibinfo{person}{Chaitanya Mishra}, \bibinfo{person}{Nick Koudas}, {and} \bibinfo{person}{Calisto Zuzarte}.} \bibinfo{year}{2008}\natexlab{}.
\newblock \showarticletitle{Generating targeted queries for database testing}. In \bibinfo{booktitle}{\emph{Proceedings of the 2008 ACM SIGMOD international conference on Management of data}}.
\newblock
\urldef\tempurl%
\url{https://doi.org/10.1145/1376616.1376668}
\showDOI{\tempurl}


\bibitem[\protect\citeauthoryear{Schmidt, Leis, Boncz, and Neumann}{Schmidt et~al\mbox{.}}{2025}]%
        {sqlstorm}
\bibfield{author}{\bibinfo{person}{Tobias Schmidt}, \bibinfo{person}{Viktor Leis}, \bibinfo{person}{Peter Boncz}, {and} \bibinfo{person}{Thomas Neumann}.} \bibinfo{year}{2025}\natexlab{}.
\newblock \showarticletitle{Sqlstorm: Taking database benchmarking into the llm era}.
\newblock \bibinfo{journal}{\emph{Proceedings of the VLDB Endowment}} \bibinfo{volume}{18}, \bibinfo{number}{11} (\bibinfo{year}{2025}), \bibinfo{pages}{4144--4157}.
\newblock


\bibitem[\protect\citeauthoryear{Seltenreich}{Seltenreich}{2020}]%
        {seltenreich2020sqlsmith}
\bibfield{author}{\bibinfo{person}{A. Seltenreich}.} \bibinfo{year}{2020}\natexlab{}.
\newblock \bibinfo{title}{{SQLsmith}}.
\newblock \bibinfo{howpublished}{\url{https://github.com/anse1/sqlsmith}}.
\newblock


\bibitem[\protect\citeauthoryear{van Renen, Horn, Pfeil, Vaidya, Dong, Narayanaswamy, Liu, Saxena, Kipf, and Kraska}{van Renen et~al\mbox{.}}{2024}]%
        {redset}
\bibfield{author}{\bibinfo{person}{Alexander van Renen}, \bibinfo{person}{Dominik Horn}, \bibinfo{person}{Pascal Pfeil}, \bibinfo{person}{Kapil Vaidya}, \bibinfo{person}{Wenjian Dong}, \bibinfo{person}{Murali Narayanaswamy}, \bibinfo{person}{Zhengchun Liu}, \bibinfo{person}{Gaurav Saxena}, \bibinfo{person}{Andreas Kipf}, {and} \bibinfo{person}{Tim Kraska}.} \bibinfo{year}{2024}\natexlab{}.
\newblock \showarticletitle{Why TPC is not enough: An analysis of the Amazon Redshift fleet}.
\newblock \bibinfo{journal}{\emph{Proceedings of the VLDB Endowment}} \bibinfo{volume}{17}, \bibinfo{number}{11} (\bibinfo{year}{2024}), \bibinfo{pages}{3694--3706}.
\newblock


\bibitem[\protect\citeauthoryear{Vuppalapati, Miron, Agarwal, Truong, Motivala, and Cruanes}{Vuppalapati et~al\mbox{.}}{2020}]%
        {snowset}
\bibfield{author}{\bibinfo{person}{Midhul Vuppalapati}, \bibinfo{person}{Justin Miron}, \bibinfo{person}{Rachit Agarwal}, \bibinfo{person}{Dan Truong}, \bibinfo{person}{Ashish Motivala}, {and} \bibinfo{person}{Thierry Cruanes}.} \bibinfo{year}{2020}\natexlab{}.
\newblock \showarticletitle{Building an elastic query engine on disaggregated storage}. In \bibinfo{booktitle}{\emph{17th USENIX Symposium on Networked Systems Design and Implementation (NSDI 20)}}. \bibinfo{pages}{449--462}.
\newblock


\bibitem[\protect\citeauthoryear{Woltmann, Hartmann, Thiele, Habich, and Lehner}{Woltmann et~al\mbox{.}}{2019}]%
        {woltmann2019cardinality}
\bibfield{author}{\bibinfo{person}{Lucas Woltmann}, \bibinfo{person}{Claudio Hartmann}, \bibinfo{person}{Maik Thiele}, \bibinfo{person}{Dirk Habich}, {and} \bibinfo{person}{Wolfgang Lehner}.} \bibinfo{year}{2019}\natexlab{}.
\newblock \showarticletitle{Cardinality estimation with local deep learning models}. In \bibinfo{booktitle}{\emph{Proceedings of the second international workshop on exploiting artificial intelligence techniques for data management}}. \bibinfo{pages}{1--8}.
\newblock


\bibitem[\protect\citeauthoryear{Yang, Kamsetty, Luan, Liang, Duan, Chen, and Stoica}{Yang et~al\mbox{.}}{2020}]%
        {yang2020neurocard}
\bibfield{author}{\bibinfo{person}{Zongheng Yang}, \bibinfo{person}{Amog Kamsetty}, \bibinfo{person}{Sifei Luan}, \bibinfo{person}{Eric Liang}, \bibinfo{person}{Yan Duan}, \bibinfo{person}{Xi Chen}, {and} \bibinfo{person}{Ion Stoica}.} \bibinfo{year}{2020}\natexlab{}.
\newblock \showarticletitle{NeuroCard: one cardinality estimator for all tables}.
\newblock \bibinfo{journal}{\emph{arXiv preprint arXiv:2006.08109}} (\bibinfo{year}{2020}).
\newblock


\bibitem[\protect\citeauthoryear{Yu, Wu, Kossmann, Li, Markakis, Ngom, Madden, and Kraska}{Yu et~al\mbox{.}}{2024}]%
        {yu2024blueprinting}
\bibfield{author}{\bibinfo{person}{Geoffrey~X Yu}, \bibinfo{person}{Ziniu Wu}, \bibinfo{person}{Ferdi Kossmann}, \bibinfo{person}{Tianyu Li}, \bibinfo{person}{Markos Markakis}, \bibinfo{person}{Amadou Ngom}, \bibinfo{person}{Samuel Madden}, {and} \bibinfo{person}{Tim Kraska}.} \bibinfo{year}{2024}\natexlab{}.
\newblock \showarticletitle{Blueprinting the Cloud: Unifying and Automatically Optimizing Cloud Data Infrastructures with BRAD--Extended Version}.
\newblock \bibinfo{journal}{\emph{arXiv preprint arXiv:2407.15363}} (\bibinfo{year}{2024}).
\newblock


\bibitem[\protect\citeauthoryear{Yue, Peng, Cai, Zhou, Hu, Zhang, Xu, and Yang}{Yue et~al\mbox{.}}{2024}]%
        {yue2024functionality}
\bibfield{author}{\bibinfo{person}{Zhongwei Yue}, \bibinfo{person}{Shujian Peng}, \bibinfo{person}{Peng Cai}, \bibinfo{person}{Xuan Zhou}, \bibinfo{person}{Huiqi Hu}, \bibinfo{person}{Rong Zhang}, \bibinfo{person}{Quanqing Xu}, {and} \bibinfo{person}{Chuanhui Yang}.} \bibinfo{year}{2024}\natexlab{}.
\newblock \showarticletitle{Functionality-aware database tuning via multi-task learning}. In \bibinfo{booktitle}{\emph{2024 IEEE 40th International Conference on Data Engineering (ICDE)}}. IEEE, \bibinfo{pages}{83--95}.
\newblock


\bibitem[\protect\citeauthoryear{Zhang, Li, Liu, Lv, and Fan}{Zhang et~al\mbox{.}}{2025}]%
        {zhang2025cloudybench}
\bibfield{author}{\bibinfo{person}{Chao Zhang}, \bibinfo{person}{Guoliang Li}, \bibinfo{person}{Leyao Liu}, \bibinfo{person}{Tao Lv}, {and} \bibinfo{person}{Ju Fan}.} \bibinfo{year}{2025}\natexlab{}.
\newblock \showarticletitle{CloudyBench: A Testbed for A Comprehensive Evaluation of Cloud-Native Databases}. In \bibinfo{booktitle}{\emph{2025 IEEE 41st International Conference on Data Engineering (ICDE)}}. IEEE Computer Society, \bibinfo{pages}{2535--2547}.
\newblock


\bibitem[\protect\citeauthoryear{Zhang, Chai, Zhou, and Li}{Zhang et~al\mbox{.}}{2022}]%
        {Learnedsqlgen}
\bibfield{author}{\bibinfo{person}{Lixi Zhang}, \bibinfo{person}{Chengliang Chai}, \bibinfo{person}{Xuanhe Zhou}, {and} \bibinfo{person}{Guoliang Li}.} \bibinfo{year}{2022}\natexlab{}.
\newblock \showarticletitle{Learnedsqlgen: Constraint-aware sql generation using reinforcement learning}. In \bibinfo{booktitle}{\emph{Proceedings of the 2022 International Conference on Management of Data}}. \bibinfo{pages}{945--958}.
\newblock


\bibitem[\protect\citeauthoryear{Zhou, Liu, Urgaonkar, Wang, Mueller, Zhang, Zhang, Pfeil, Horn, Liu, Pagano, Kraska, Madden, and Fan}{Zhou et~al\mbox{.}}{2025}]%
        {pbench}
\bibfield{author}{\bibinfo{person}{Yan Zhou}, \bibinfo{person}{Chunwei Liu}, \bibinfo{person}{Bhuvan Urgaonkar}, \bibinfo{person}{Zhengle Wang}, \bibinfo{person}{Magnus Mueller}, \bibinfo{person}{Chao Zhang}, \bibinfo{person}{Songyue Zhang}, \bibinfo{person}{Pascal Pfeil}, \bibinfo{person}{Dominik Horn}, \bibinfo{person}{Zhengchun Liu}, \bibinfo{person}{Davide Pagano}, \bibinfo{person}{Tim Kraska}, \bibinfo{person}{Samuel Madden}, {and} \bibinfo{person}{Ju Fan}.} \bibinfo{year}{2025}\natexlab{}.
\newblock \showarticletitle{PBench: Workload Synthesizer with Real Statistics for Cloud Analytics Benchmarking}.
\newblock \bibinfo{journal}{\emph{Proc. VLDB Endow.}} \bibinfo{volume}{18}, \bibinfo{number}{11} (\bibinfo{date}{July} \bibinfo{year}{2025}), \bibinfo{pages}{3883–3895}.
\newblock
\showISSN{2150-8097}
\urldef\tempurl%
\url{https://doi.org/10.14778/3749646.3749661}
\showDOI{\tempurl}


\end{thebibliography}

\end{document}